\documentclass[%
preprint,
 amsmath,amssymb,
 aps,
]{revtex4-2}

\usepackage{graphicx}
\usepackage{dcolumn}
\usepackage{bm}

\begin{document}


\title{Quantum Control of Thermal Emission from Photonic Crystals with Two-Level Atoms}

\author{Chih-Wei Wang} 
\author{Jhih-Sheng Wu}
\email{e-mail address: jwu@nycu.edu.tw}

\address{Department of Photonics, College of Electrical and Computer Engineering, National Yang Ming Chiao Tung University, Hsinchu 30010, Taiwan}

\date{\today}

\begin{abstract}

Thermal light engineering is a field of considerable interest and potential. We study quantum light-matter interactions in a one-dimensional photonic crystal with two-level atoms as the active medium, replacing classical oscillators in traditional blackbody models. In a thermal bath with pumping, these atoms modulate thermal emission via interactions with photonic modes. The model with quantum two-level systems enables the processes of spontaneous emission, stimulated absorption, and stimulated emission.
Equilibrium and nonequilibrium regimes depend on competition between pumping and thermal relaxation rates. Strong light-matter interaction and photon decay govern dynamics and steady states. 
In equilibrium, with a high thermal relaxation rate, photon numbers are initially determined by spontaneous emission and later stabilize due to stimulated absorption, influenced by light-matter interaction strength.
In-band-gap photons reach steady states at a time scale of one or two orders of magnitude longer than outside-band-gap photons. Interestingly, for a strong light-matter interaction, all photons in the equilibrium regimes show Planckian radiation, regardless of their frequencies in or out of the band gaps. Band-gap suppression of thermal emission is more pronounced with weaker light-matter interaction or larger photon decay. In the nonequilibrium regime, the dynamics of photon numbers exhibit a multi-time-scale process transitioning to steady states due to strong pumping and stimulated processes. Steady-state electron populations of two-level atoms deviate from the Fermi-Dirac distribution, and the steady-state photon numbers exhibit super-Planckian emission. 
These findings enable quantum control of thermal emission spectra, which is relevant for reducing thermal noise in quantum computing or enhancing radiative cooling.

\end{abstract}

\maketitle


\section{\label{sec:level1}Introduction}

Thermal radiation, one of the foundational topics in quantum physics, is exemplified by the Planckian radiation emitted by a blackbody. This phenomenon arises from photons with quantized energies in thermal equilibrium with a blackbody, typically modeled as a closed cavity with a small aperture. The cavity has to be sufficiently large to match the photon density of states in free space, at least for the frequencies under consideration. In traditional derivations, photons across all frequencies are assumed to fully interact with the cavity walls \cite{LandauLifshitz1980}. In this framework, photons are treated quantum-mechanically, while the matter comprising the cavity walls is modeled as classical oscillators spanning all frequencies.
Advances in the study of thermal radiation have expanded this model by relaxing key assumptions, including those related to photon density of states, thermal equilibrium, the scattering properties of matter, and the classical oscillator model of matter.

Nanophotonic structures have revolutionized the study of thermal radiation by enabling precise control over the optical photon density of states (DOS) and scattering properties \cite{baranov2019nanophotonic}. Photonic crystals can tailor the DOS to selectively enhance or suppress thermal emission at specific wavelengths \cite{Cornelius1999, Lin2000, Lin2003, Narayanaswamy2004, Luo2004,han2007tailoring}. Plasmonic nanostructures, which allow strong field confinement, exhibit high emissivity, directional emission, and efficient near-field radiative heat transfer \cite{kravets2008plasmonic, Costantini2015, aydin2011broadband, Kittel2005, Ben-Abdallah2008}. Metasurfaces, capable of multifunctional light modulation, have recently been employed to control the narrow-band polarization of thermal radiation \cite{nolen2024local}. Nanophotonics offers a versatile platform for manipulating thermal radiation beyond the constraints of classical blackbody models \cite{li2018nanophotonic}.

Thermal radiation in a nonequilibrium and active regime has gained significant attention for exploring non-Planckian radiation. 
Systems under electrical bias or optical excitation can develop nonequilibrium electron and photon populations.
In nonequilibrium scenarios, the photon number deviates from the Planckian radiation, leading to novel emission characteristics \cite{wurfel1982chemical}.  
A recent study has demonstrated a nonlinear relationship between light emission and electric bias power \cite{hsieh2025observation}. This observed nonlinear behavior departs from the classical oscillator model for matter, highlighting the need for a quantum model.

We shift focus to the quantum regime of matter, motivated by the limitations of classical oscillator models in describing thermal emission in advanced photonic systems. While prior studies have primarily treated matter as classical oscillators, recent advances in quantum optics suggest that incorporating quantum mechanical descriptions of matter, such as two-level atoms, can unlock new regimes of control over thermal radiation \cite{ridolfo2013nonclassical,pirkkalainen2015cavity,askenazi2017midinfrared,Pilar2020thermodynamicsof}. The theoretical work by Chow examines thermal emission using two-level atoms as an active medium \cite{chow2006theory}. This model employs a fully quantum approach, allowing for a comprehensive description of complicated competition processes. The study demonstrated the conditions necessary for nonequilibrium and super-Planckian radiation, focusing on pumping and thermal relaxation rates. 
Nonequilibrium electron populations occur when the pumping rate is comparable to the thermal relaxation rate, while a fast thermal relaxation rate leads to equilibrium electron populations.
 However, the effects of light-matter interaction and photon decay have not yet been thoroughly investigated. In a photonic environment, the variations in field confinement and photon loss can significantly impact radiation characteristics. 
 Understanding these effects is crucial in this area.

\begin{figure}
\includegraphics[width=8.5cm]{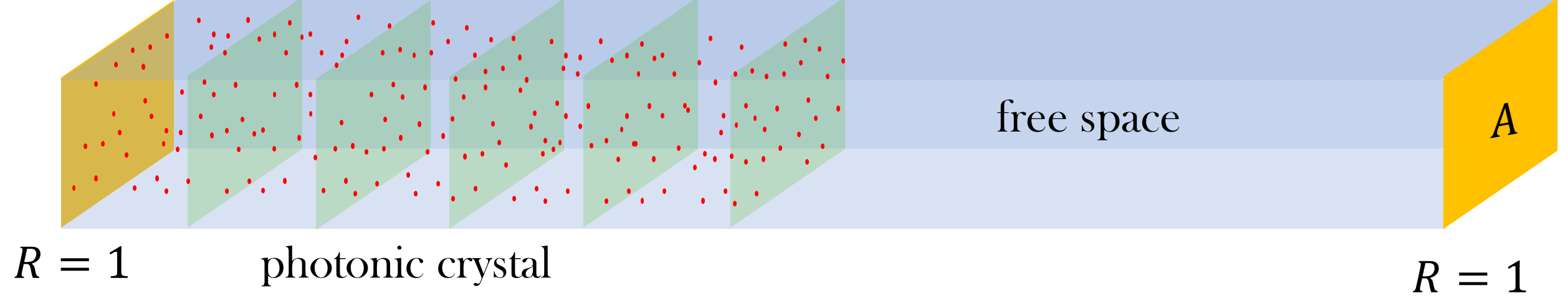}
\caption{\label{fig:sch} Schematic of a one-dimensional photonic crystal within a large cavity simulating free space. The system consists of the left photonic crystal and the surrounding free space, which is significantly larger than depicted in the schematic. The system has the perfect boundary conditions, where the reflections are perfect at $z=0$ and $z=L$. The two-level atoms, acting as an active medium, are uniformly distributed in the photonic crystal. The area shaded in blue represents the space with a free-space permittivity $\epsilon_0$.  The photonic crystal is modeled with periodic semi-transparent planes. Each plane (shaded in green) is modeled by a delta function in permittivity. 
}
\end{figure}

In this research, we examined how light-matter interaction influences both the dynamic and steady-state properties of thermal emissions. 
Our findings indicate that the strength of light-matter interaction establishes the upper limit of thermal emissions, especially for in-band-gap photons. The strength of light-matter interaction is proportional to the density: 
\begin{align}
n_{\mathrm{atom}} = \frac{N_j}{A L_c},
\end{align}
where $N_j$ is the number of atoms per frequency $\omega_n$  (as defined in Sec.~\ref{sec:qelm}),
 and $A$ and $L_c$ are the area and the length of the photonic crystal, respectively.
 A larger $n_{\mathrm{atom}}$ can be achieved by a smaller volume or a larger $N_j$. 
Strong field confinement (field intensity in the active medium) can also enhance light-matter interaction. 
Previous studies have shown that the emission of in-band-gap photons is generally suppressed \cite{Luo2004, chow2006theory}. However, we find that steady-state emission of in-band-gap photons is not suppressed in the case of strong light-matter interaction.  
In-band-gap photons have small field confinement in photonic crystals, resulting in slower dynamics than outside-band-gap photons.
Interestingly, in the case of strong light-matter interaction, the photon numbers of in-band-gap photons reach steady states without band-gap suppression at a time scale that is one or two orders of magnitude longer than outside-band-gap photons.
Furthermore, our results indicate that large photon loss can significantly suppress photon numbers in band gaps. In our model, photon loss encompasses both material and radiative losses. The degree of band-gap suppression exhibits a positive correlation with the amount of photon loss. These findings highlight the complex interactions affecting thermal radiation. These interactions are evident in the multi-time-scale dynamics of photon numbers in both the equilibrium and non-equilibrium regimes. The origins of multi-time-scale dynamics stem from competition among thermal relaxation, pumping, photon loss,  and spontaneous and stimulated processes. 
Our findings clarify how light-matter interaction and photon loss influence the dynamics and steady states of active thermal radiation in photonic crystals.

A schematic of the considered model is shown in Fig.~\ref{fig:sch}. The closed system, consisting of a one-dimensional (1D) photonic crystal and free space, simulates the universe. This approach has been implemented to address the theoretical issue concerning the outcoupling of photonic crystal eigenmodes into free space \cite{Lang1973,chow2006theory}. With this approach, the photonic eigenmodes of the system inherently reflect the interactions between the photonic crystal and free space.
 The system has a total length $L$, and the photonic crystal is positioned from $z=0$ to $z=L_c$. The perfect boundary conditions, the reflection $R=1$, are imposed at $z=0$ and $z=L$. The length $L$ should be much larger than $L_c$. We use $L=1.2$~cm and $L_c=120~\mu$m.  The active medium consists of the two-level atoms, uniformly distributed in the photonic crystal. We model a one-dimensional crystal composed of periodic semitransparent planes for brevity, as in Refs.~\onlinecite{Lang1973,chow2006theory}. The photonic crystal consists of 12 semitransparent planes, with a lattice constant of $l_p=10~\mu$m. The 1D system has an effective area $A$, which determines light-matter interaction strength.   
 
 We describe the eigenmodes of the whole system in Sec.~\ref{sec:mode}. The modes are categorized into photonic and in-band-gap modes based on their field distribution.
The quantum model of photons, two-level atoms, and light-matter interaction is described in Sec.~\ref{sec:qelm}. Following the established methods, we present the derivation of the quantum dynamic equations for photon numbers and electron populations for clarity and completeness. Our main results from the dynamic equations are discussed in Secs.~\ref{sec:dy} and ~\ref{sec:ss}. 
In Section~\ref{sec:dy}, we solve the dynamical equations directly using the Runge-Kutta method. The competition natures are reflected in the multi-time-scale dynamics of photon numbers.  The mechanisms of each timescale are analyzed. The results demonstrate the competition among various processes: stimulated excitation and emission, thermal relaxation, pumping, and photon loss.
In Sec.~\ref{sec:ss}, we obtain the steady-state solutions with the Newton-Krylov method. We show and discuss how free-space thermal emission depends on light-matter interaction and photon loss.
Concluding remarks and future outlooks based on our results are discussed in Sec.~\ref{sec:conc}.

\section{Eigenmodes  of the System}\label{sec:mode}
We consider a one-dimensional closed system to simulate the thermal radiation of a photonic crystal in free space, as shown in Fig.~\ref{fig:sch}. This one-dimensional model effectively captures the main physics observed in three-dimensional calculations \cite{chow2006theory}. The system has the perfect reflections $R=1$ at the boundaries, $z=0$, and $z=L$. Most space has the vacuum permittivity $\epsilon_0$, while twelve semitransparent planes  are located at $z = n_{\mathrm{st}}l_p$, with $n_{\mathrm{st}} = 1,\,2,\,3,\,\ldots,\,12$. The periodic semitransparent planes serve as a simplified model for a photonic crystal, exhibiting features such as band gaps and dispersion. The eigenmodes of the whole system should be interpreted as a hybridization of free-space plane waves and modes of the photonic crystal. Such eigenmodes inherently contain the interaction between photons of photonic crystals and free space. This approach avoids the issue of determining the outcoupling of modes of a photonic crystal to free space \cite{Lang1973}. 

\begin{figure}
\includegraphics[width=5.5cm]{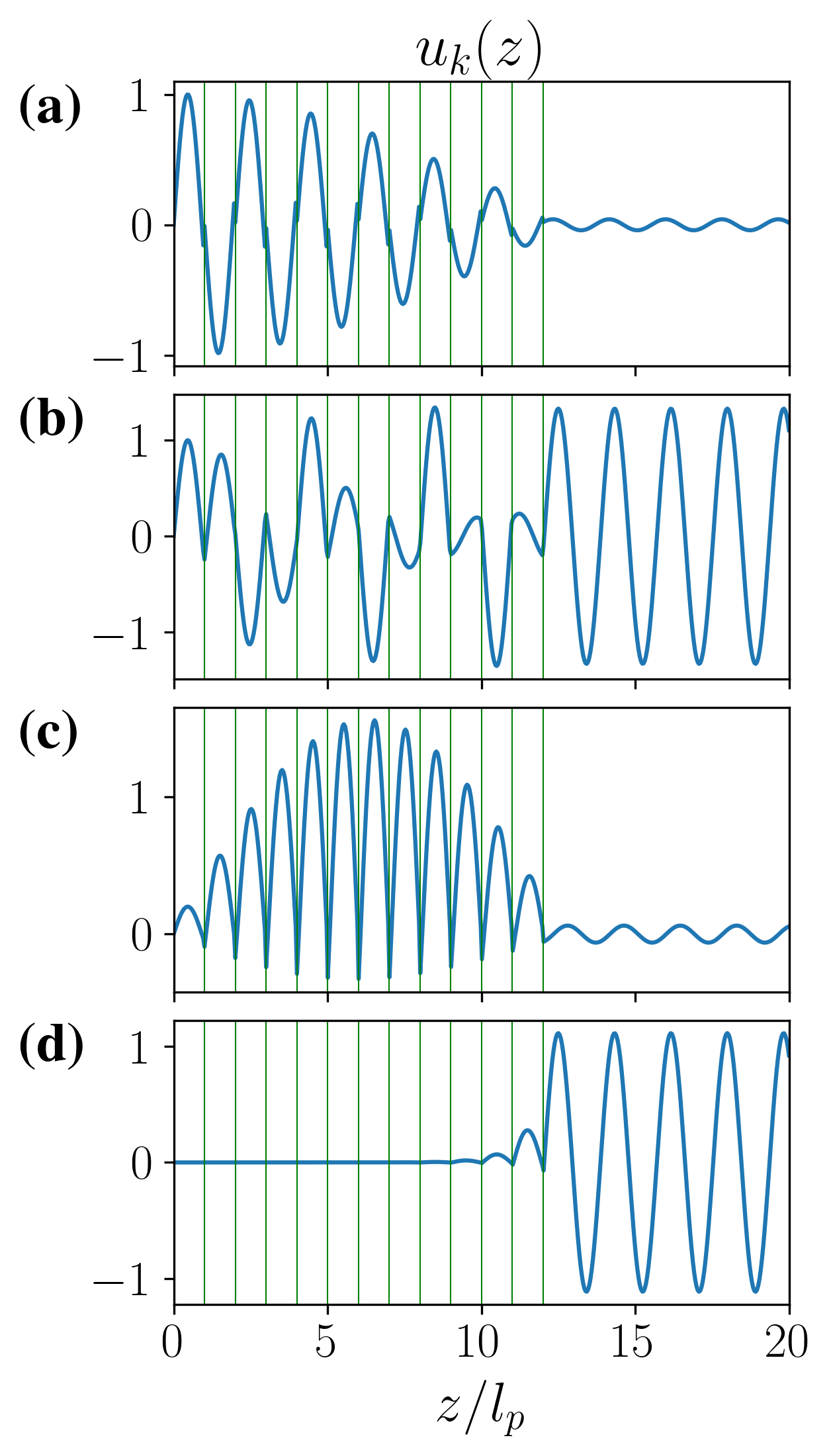}
\caption{\label{fig:modes} 
 Eigenfunctions $u_{k}(z)$ in arbitrary units for $\eta=2.1\times10^{-5}$~m. The corresponding band structure is shown in Fig.~\ref{fig:band}c.
 (a) the band-edge mode of $\Omega_k =9.43\times10^{13}~\frac{\mathrm{rad}}{\mathrm{s}}$ and $\Gamma_k =0.76$,
(b) the band-center mode of $\Omega_k =1.02\times10^{14}~\frac{\mathrm{rad}}{\mathrm{s}}$ and $\Gamma_k =4.8\times10^{-3}$,
 (c) the band-edge mode of $\Omega_k =1.09\times10^{14}~\frac{\mathrm{rad}}{\mathrm{s}}$ and $\Gamma_k =0.61$,
 and (d) the in-band-gap mode of $\Omega_k =1.17\times10^{14}~\frac{\mathrm{rad}}{\mathrm{s}}$ and $\Gamma_k =4.9\times10^{-5}$.
}
\end{figure}

\begin{figure}
\includegraphics[width=9cm]{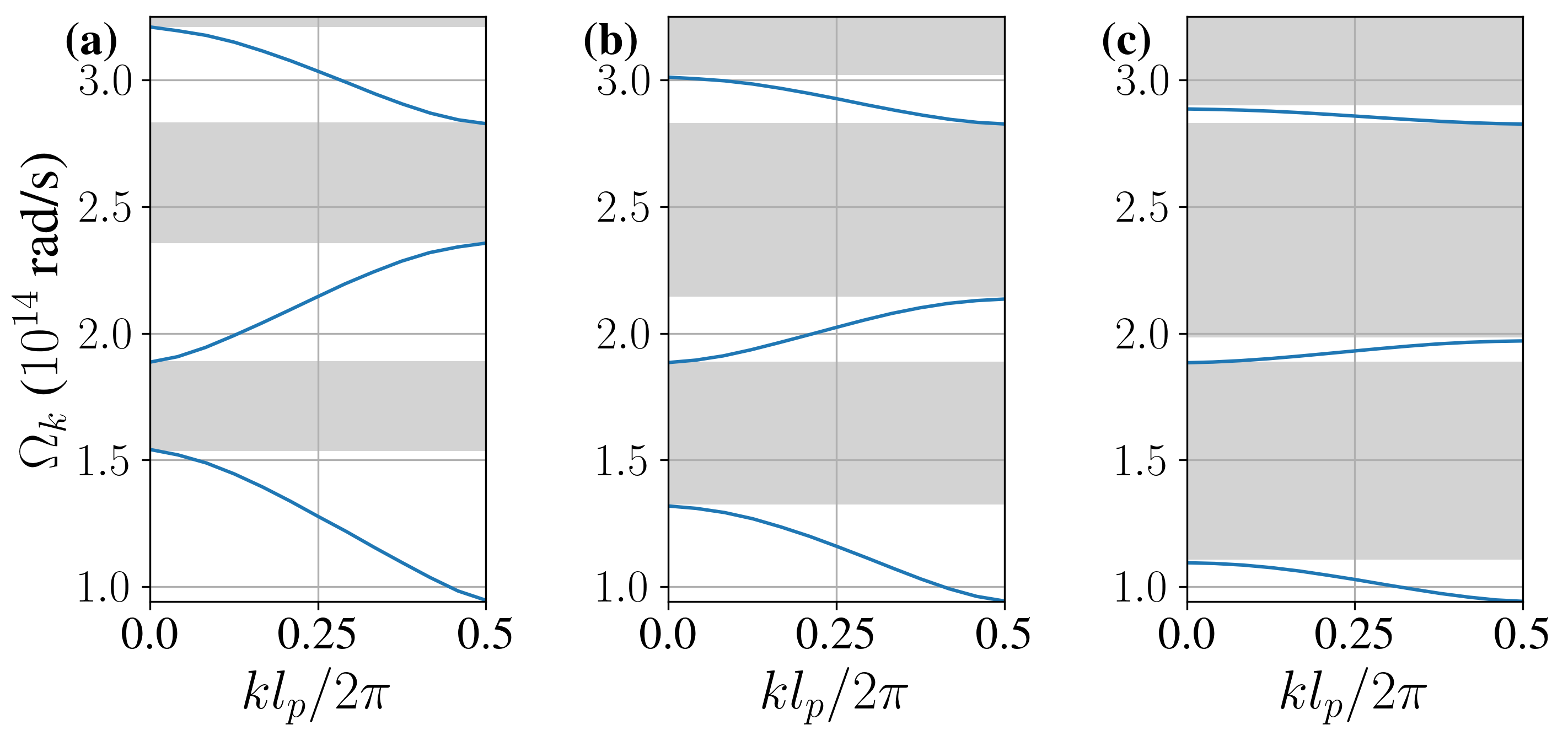}
\caption{\label{fig:band} 
Band structures of the modes of the photonic crystals with (a) $\eta =2.6\times10^{-6}$~m, (b) $\eta =6.4\times10^{-6}$~m, and (c) $\eta =2.1\times10^{-5}$~m, respectively. The complete band gaps are shaded in gray.
}
\end{figure}

The electric field of an eigenmode $\mathbf{E}_k(z)  = \sqrt{\frac{\hbar\Omega_k}{\epsilon_0 A}}u_k(z) \hat{\mathbf e}$ in the $i$th region can be written as 
\begin{align}
 \mathbf{E}_k(z) &= \sqrt{\frac{\hbar\Omega_k}{\epsilon_0 A}}\left(a_i e^{i \frac{\Omega_k}{c} z} + b_i e^{-i\frac{\Omega_k}{c} z}\right)\hat{\mathbf e},\label{eq:Ek}
\end{align}
where $k$ is the label of the eigenmode. The $i$th region is at $z_i<z<z_{i+1}$, with $z_i$ being the positions of the boundaries and the semitransparent planes.  The permittivity of the system is
\begin{align}
\epsilon(z) = \epsilon_0\left[1 + \eta \sum_{i=1}^{12}\delta(z-z_i)\right],
\end{align}
where the Dirac delta functions model the semitransparent planes, and 
$\eta$ is a strength constant with the unit of length.
By matching the electric field of Eq.~\eqref{eq:Ek} at $z_i$ with the boundary conditions and the transfer matrix method, all the coefficients $a_i$ and $b_i$ can be eliminated, and an equation of the eigenfrequency $\Omega_k$ is obtained.  We numerically solve the equation in search of eigenfrequencies $0<\Omega_k<~5\times10^{14}~\frac{\mathrm{rad}}{\mathrm{s}}$ and the corresponding coefficients, $a_i$ and $b_i$. 
For the energy quantization of each mode,
 \begin{align}
 \int \mathbf{D}_k\cdot\mathbf{E}_kdv = \hbar \Omega_k,
\end{align} 
the eigenfunctions $u_k(z)$ satisfy the normalization condition,
\begin{align}
\int_0^L \epsilon(z)|u_k(z)|^2 dz = \epsilon_0,
\end{align}
which fixes all the coefficients $a_i$ and $b_i$.

The system has approximately 7000 eigenfunctions in the considered frequency range ($0<\Omega_k<~5\times10^{14}~\frac{\mathrm{rad}}{\mathrm{s}}$).  These modes are categorized into photonic crystal modes and in-band-gap photons based on the field confinement factor 
\begin{align}
\Gamma_k =\int_0^{L_c} |u_k(z)|^2 dz.
\end{align}
The modes with $\Gamma_k > \frac{L_c}{L}$ are categorized into photonic crystal modes, while the others are in-band-gap photons. 

 Figure~\ref{fig:modes} shows the eigenmodes of the photonic crystal of  $\eta=2.1\times10^{-5}$~m, of which band structure is shown in Fig.~\ref{fig:band}c.  These modes are categorized by $\Gamma_k$, where modes outside the band gap have greater $\Gamma_k$, while in-band-gap modes have very small $\Gamma_k$.
Note that in the previous step, $k$ is only a label of an eigenfunction. For these photonic crystal modes, each $k$ can be assigned with the physical value $2\pi m_k^{(\mathrm{peak})}/l_c$, where $ m_k^{(\mathrm{peak})}$ is the number of peaks of $u_k(z)$ in the photonic crystal. With these $k$ and $\Omega_k$, the band structures can be plotted as shown in Fig.~\ref{fig:band}. 
The frequencies of Figs.~\ref{fig:modes}a, \ref{fig:modes}b,
 and \ref{fig:modes}c are at the bottom band edge, the band center, and the top band edge of Fig.~\ref{fig:band}c, while
 the frequency of Fig.~\ref{fig:modes}d is in the band gap. Fig.~\ref{fig:modes}d shows an
in-band-gap photon, which has a small field confinement $\Gamma_k$. In-band-gap photons exhibit small penetration lengths. 

Photonic band gaps were shown to suppress thermal emissions \cite{Narayanaswamy2004,
Luo2004, chow2006theory}. Our findings, however, indicate that the suppression associated with band gaps can diminish in the presence of strong light-matter interactions, even under equilibrium conditions, which is discussed in Sec.~\ref{sec:ss}. The origin is that in-band-gap photons possess a non-zero electric field within the photonic crystal. Such small field confinements $\Gamma_k$ can still lead to Planckian radiation over a longer time scale of dynamics, as discussed in Sec.~\ref{sec:dy}.

\section{Quantum Equations of Light-Matter Interaction}\label{sec:qelm}

The electric field operators of  each eigenmode in the Heisenberg picture are \cite{gerry2023introductory}
\begin{align}
    \mathbf{E}_{k}(z,t) =  \sqrt{\frac{\hbar\Omega_k}{\epsilon_0 A}}\left[ a_k(t) + a_k^\dagger(t)\right]u_k(z)\hat{\mathbf e}\,,
\end{align}
where $ a^\dagger_k$ and $a_k$ are the photon creation and annihilation operators. 
The two-level atoms are assumed to be uniformly distributed in the photonic crystal. 
Each atom, labeled by indices $n$ and $j$, is located at a position $z_j$ and has a resonant frequency $\omega_n$.
The number of the position indices is $N_j$. Since the atoms are uniformly distributed, we define the density of the atoms to be 
$n_{\mathrm{atom}} = N_j/(AL_c)$.
We select a uniformly distributed frequency, denoted as $ \omega_n$, within the desired range, ensuring that $ 0 < \omega_n < 5\times10^{14}~\frac{\mathrm{rad}}{\mathrm{s}}$.
The number of frequencies $N_\omega$ is set as 500, approximating a continuous frequency distribution.
The ground and excited states of the atom of the indices $n$ and $j$ are represented by $|g_{nj}\rangle$ and $|e_{nj}\rangle$, respectively.   The Hamiltonian of the atoms and fields in the rotating-wave approximation is
\begin{align}
    H &=  \sum_k\,\hbar\Omega_k a_k^\dagger a_k +
\sum_{n,j}\,\hbar \omega_n |e_{nj}\rangle \langle e_{nj}| \notag \\    &- \sum_{k,n,j} g_{kj}\left( a_k\sigma_{nj}^{\dagger}  + a^\dagger_k \sigma_{nj} \right)\,,
\end{align}
where $g_{kj}=\mu \sqrt{\frac{\hbar\Omega_k}{\epsilon_0 A}} u_k(z_j)$ is the interaction strength, $\mu$ is the dipole matrix element of the atom, and $\sigma_{nj}=|g_{nj}\rangle\langle e_{nj}|$ is the lowering operator.  

The equations of motion for the physical quantity operators,  derived from the Heisenberg equation, are
\begin{align}
\frac{d p_{njk}}{d t} &= \frac{i}{\hbar}e^{-i\omega_{nk}t}\sum_{k'}\,g_{k'j}(\hat{n}_{nj}^{(e)} a_ka_{k'}^\dagger - \hat{n}_{nj}^{(g)} a_{k'}^\dagger a_k),\label{eq:p}\\
\frac{d \hat{n}_{nj}^{(e)}}{d t} &= -\frac{i}{\hbar}\sum_{k'}\,g_{k'j}(p_{njk}^{\dagger}e^{-i \omega_{nk}t} - p_{njk}e^{i\omega_{nk}t}),\label{eq:n}\\
\frac{d \hat{N}_k}{d t} &= \frac{i}{\hbar}\sum_{n,j}\,g_{k'j}(p_{njk}^{\dagger}e^{-i \omega_{nk}t} - p_{njk}e^{i \omega_{nk}t}),\label{eq:Nk}
\end{align}
where $p_{njk}$ is the polarization defined by 
\begin{align}
    p_{njk}= \sigma_{nj}^{\dagger}a_k\exp\left(-i\omega_{nk}t\right).
\end{align}
$\omega_{nk}=\omega_n-\Omega_k$ is the detuning.
$\hat{n}^{(e)}_{nk} = |e_{nj}\rangle\langle e_{nj}|$ and $\hat{n}^{(g)}_{nk} = |g_{nj}\rangle\langle g_{nj}|$ are the population operators of the excited and ground states, respectively. Note $\hat{n}^{(g)}_{nk}+\hat{n}^{(e)}_{nk}=  \mathbb{I}$ for conservation of electrons. $\hat{N}_k=a^{\dagger}_ka_k$ is the photon number operator.
To describe the decays and losses of the system, the decay constants are added as phenomenological parameters in the dynamical equations.
 Due to the dephasing collisions, the dynamical equation of the polarization with a dephasing rate $\gamma$ becomes
\begin{align}
\frac{d p_{njk}}{d t} =& \frac{i}{\hbar}e^{-i\omega_{nk}t}\sum_{k'}\,g_{k'j}(\hat{n}_{nj}^{(e)} a_ka_{k'}^\dagger - \hat{n}_{nj}^{(g)} a_{k'}^\dagger a_k)\notag\\
&-\gamma p_{njk}.
\end{align}

Assuming the effective dephasing rate $\gamma$ is much faster than the rate of changes of other operators, the polarization ${p}_{njk}$ can be treated adiabatically as follows:
\begin{align}
    {p}_{njk} e^{i\omega_{nk}t}= \frac{i}{\hbar}\sum_{k'}\frac{g_{k'j}}{\gamma-i\omega_{nk}}\left(\hat{n}^{(e)}_{nj} a_k a_{k'}^\dagger - \hat{n}^{(g)}_{nj}a_{k'}^\dagger a_k \right).
\end{align}
Taking the expectation values  on  both sides, we obtain
\begin{align}
\langle {p}_{njk}  \rangle  e^{i\omega_{nk}t } &= \frac{ig_{kj}}{\hbar(\gamma-i\omega_{nk})}\times\notag\\
&\left[ \langle \hat{n}^{(e)}_{nj} \rangle \left(\langle \hat{N}_k \rangle +1\right) - \langle \hat{n}^{(g)}_{nj} \rangle \langle \hat{N}_k  \rangle \right],
\label{eq:p_expect}
\end{align}
where we have utilized the property that $\langle a_k a_{k'}^\dagger \rangle = 0$ if $k\neq k'$. Taking the expectation values of the other operators in
Eqs.~\eqref{eq:n} and \eqref{eq:Nk} and substituting $\langle p_{njk} \rangle$ with Eq.~\eqref{eq:p_expect}, we obtain
\begin{align}
\frac{d n^{(e)}_{nj}}{d t} &= - \sum_k\frac{2g_{kj}^2}{\hbar^2\gamma} \left[(2n^{(e)}_{nj}-1)N_k +n^{(e)}_{nj} \right]L_{nk},\label{eq:dne_1}
 \\
\frac{d N_k}{d t} &= \sum_{n,j}\frac{2g_{kj}^2}{\hbar^2\gamma} \left[(2n^{(e)}_{nj}-1)N_k +n^{(e)}_{nj} \right]L_{nk}, \label{eq:dNk_1}
\end{align}
with $L_{nk}=[1+(\omega_n-\Omega_k)^2/\gamma^2]^{-1}$,   $n^{(e)}_{nj} = \langle \hat{n}^{(e)}_{nj} \rangle$, and 
$N_k = \langle \hat{N}_{k} \rangle$. We have used the property that $n^{(e)}_{nj} + n^{(g)}_{nj}=1$.

Since the atoms are uniformly distributed, we define the average population $n^{(e)}_{n}$ by 
\begin{align}
n_{n}^{(e)} = \frac{\sum_{_j}n_{nj}^{(e)}}{N_j} \label{eq:ne_ave}.
\end{align} 
To make the computation of the light-matter interaction within the photonic crystal tractable, we adopt an effective medium approximation, following the approach outlined in \cite{chow2006theory}. Specifically, we replace the local field intensity $|u_k(z_j)|^2$ at discrete lattice sites $j$ with its spatial average over the photonic crystal. This leads to the approximation
\begin{align}
\sum_{j}|u_k(z_j)|^2 n^{(e)}_{nj} \sim n_{n}^{(e)} \frac{\Gamma_k}{L_c}, \label{eq:em_app}
\end{align}
with 
\begin{align}
\Gamma_k=\int_0^{L_c} |u_k(z)|^2 dz \label{eq:Gamma_k}
\end{align}
being the mode confinement factor.  Using Eqs.~\eqref{eq:ne_ave} and \eqref{eq:em_app} to eliminate $j$ in 
 Eqs.~\eqref{eq:dne_1} and \eqref{eq:dNk_1}, the primary equations for the main results of this study are obtained:
\begin{align}
\frac{d n^{(e)}_{n}}{d t} &= - \frac{2\mu^2n_{V}}{\hbar\epsilon_0 \gamma}\sum_k \Omega_k\Gamma_k\left[(2n^{(e)}_{n}-1)N_k +n^{(e)}_{n} \right]L_{nk} \notag\\
&- \gamma_{r}\left[n^{(e)}_n - f_n(T)\right] + \Lambda(\omega_n,
T) (1 - n_{n}^{(e)}),\label{eq:dne_f}
 \\
\frac{d N_k}{d t} & =\frac{2\mu^2 n_{\mathrm{atom}}}{\hbar\epsilon_0 \gamma} \sum_{n} \Omega_k\Gamma_k\left[(2n^{(e)}_{n}-1)N_k +n^{(e)}_{n} \right]L_{nk} \notag\\
&-\gamma_c N_k, \label{eq:dNk_f}
\end{align}
where 
\begin{align}
f_n (T) = \frac{1}{1 +\exp\left(\frac{\hbar\omega_n}{k_BT}\right)}
\end{align}
 is the Fermi-Dirac distribution, $T$ is the reservoir temperature,  and $n_V =\frac{1}{AL_c}$. The thermal relaxation rate $\gamma_r$ is added to describe the thermal relaxation of electrons. The photon decay rate, $\gamma_c$, models both radiative and non-radiative losses. 
 A  phenomenological temperature-dependent pumping rate, denoted as $\Lambda(\omega_n, T)$, is introduced to model the effects of temperature on pumping. 
 The pumping rate $\Lambda(\omega_n,T)$ is
\begin{align}
\label{eq:pumping_rate}
    \Lambda(\omega_n,T) ~=~ \Lambda_0 \exp\left[ \frac{\hbar(\omega_0 - \omega_n)}{k_B T} \right]\,,
\end{align}
where $\Lambda_0$ is the pumping strength,  and $\hbar\omega_0$ is the material band-gap energy.

The  thermal emission spectrum  in free space is described by 
\begin{align}
S_\mathrm{out}(\omega) =\Omega_k N_k (1-\Gamma_k)|_{\Omega_k=\omega}, 
\end{align}
which is the photon energy outside the photonic crystal. Taking the frequency resolution $\gamma_d$ of a detector into account, the thermal emission spectrum measured by a detector is defined by 
\begin{align}
\tilde{S}_\mathrm{out}(\omega) =\sum_k \frac{1}{1+\left(\frac{\omega-\Omega_k}{\gamma_d}\right)^2}\Omega_k N_k (1-\Gamma_k), \label{eq:Sout}
\end{align}
where the Lorentzian function describes the finite frequency resolution $\gamma_d$ of a detector.

Thermal emission can be classified into two regimes based on electron populations: equilibrium and nonequilibrium \cite{chow2006theory}.
When thermal relaxation occurs much faster than pumping ($\gamma_r \gg \Lambda_0$), electron populations adhere to the Fermi-Dirac distribution. In contrast, when thermal relaxation is similar to pumping, electron populations begin to deviate from the Fermi-Dirac distribution. In the following sections, we discuss the dynamics and steady-state solutions according to the equilibrium and nonequilibrium regimes.

\section{Dynamics of Photon Numbers} \label{sec:dy}

The dynamics of photons and electrons can reveal the dominant mechanism of light-matter interaction. In this quantum model,
photon dynamics can be asynchronous with electron dynamics, resulting in multi-time-scale behaviors. The asynchronous behavior of photons and electrons stems from the strength of light-matter interaction and the loss of photons. 
The strength of light-matter interaction is determined by $n_\mathrm{atom}$ and $\Gamma_k$. 
We examine the effects of $n_\mathrm{atom}$ , $\Gamma_k$, and $\gamma_c$ on the dynamics.

We computed the dynamics of the photon numbers and populations by numerically solving  Eqs.~\eqref{eq:dne_f} and ~\eqref{eq:dNk_f}  with the Runge-Kutta method. In total, a system of 7700 differential equations is computed, where 7200 photonic modes and $N_\omega=500$ atomic levels are taken into account. We set all the $N_k(t=0)=0$ and $n^{(e)}_n(t=0)=0$. The simulation parameters are listed in Table~\ref{tab:parameters}, except for those specified in each case.

\begin{figure*}
\includegraphics[width=11.5cm]{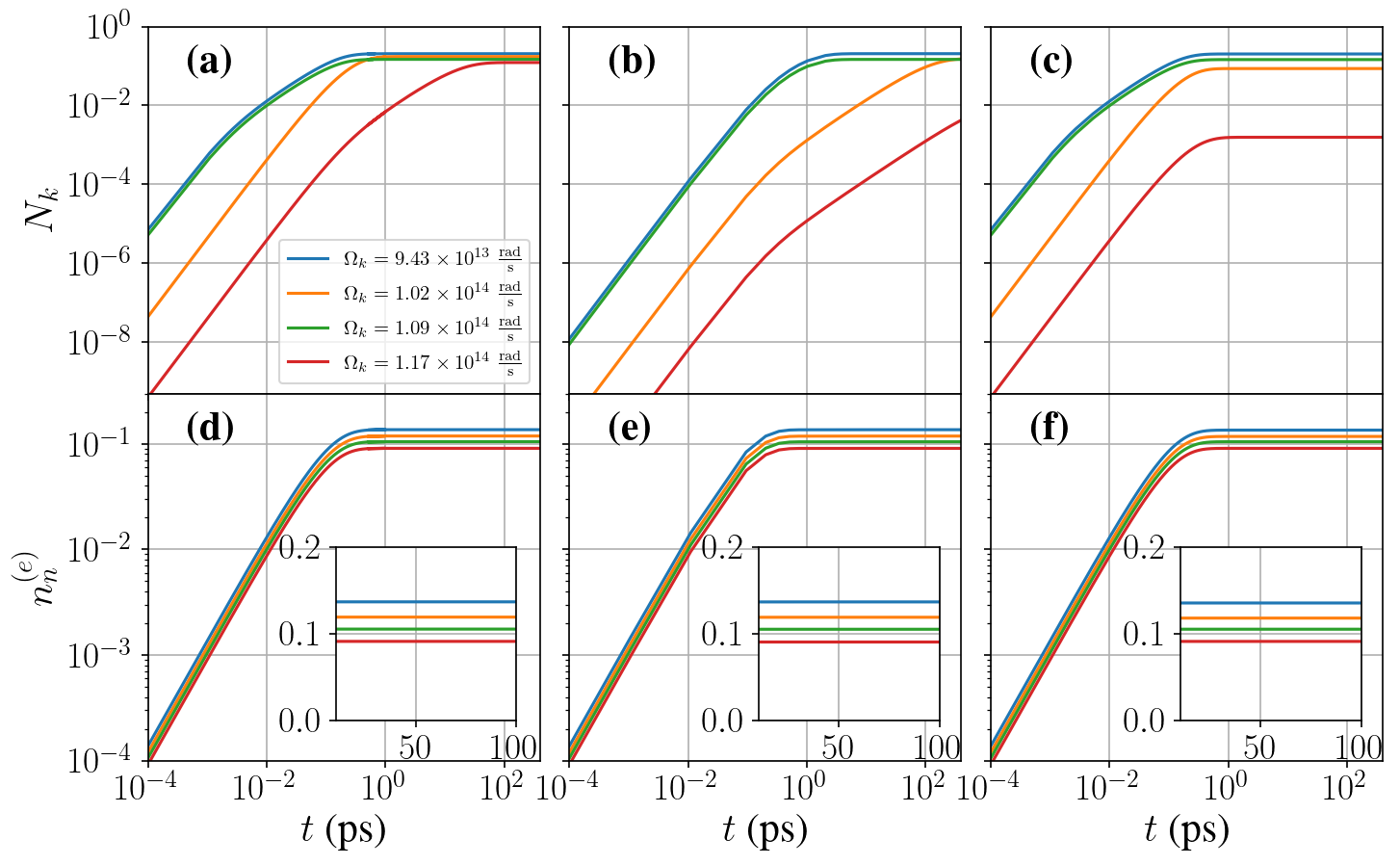}
\caption{\label{fig:dy_com_eq} Dynamics of photons and electrons in the equilibrium regime. Panels (a) and (d) correspond to Case (i): $n_\mathrm{atom}=5\times10^{24}~\mathrm{m}^{-3}$ and  $\gamma_c=10^{9}~\mathrm{s}^{-1}$.  Panels (b) and (e) correspond to Case (ii): $n_\mathrm{atom}=5\times10^{22}~\mathrm{m}^{-3}$ and  $\gamma_c=10^{9}~\mathrm{s}^{-1}$. Panels (c) and (f) correspond to Case (iii): $n_\mathrm{atom}=5\times10^{24}~\mathrm{m}^{-3}$ and  $\gamma_c=5\times10^{12}~\mathrm{s}^{-1}$. The electron populations $n^{(e)}_n(\omega)$ of $\omega = \Omega_k$ are plotted.  
The insets in (d), (e), and (f) show the steady-state electron populations at a large $t$, which closely follow the Fermi-Dirac distribution. 
}
\end{figure*}
\subsection{Equilibrium}
For the equilibrium regime, we set the thermal relaxation rate  $\gamma_r = 10^{13}~\mathrm{s}^{-1}$ and the pumping strength $\Lambda_0= 10^{10}~\mathrm{s}^{-1}$, at which the electron population $n^{(e)}_n$ can always saturate to the Fermi-Dirac distribution.   Figure~\ref{fig:dy_com_eq} shows the dynamics of electrons and photons in a photonic crystal with a parameter of $\eta = 2.1\times 10^{-5}$~m, as this photonic crystal exhibits more pronounced band gaps (Fig.~\ref{fig:band}).
We consider three cases to analyze the effects of the strength of light-matter interaction and photon loss: (i) a large $n_\mathrm{atom}=5\times10^{24}~\mathrm{m}^{-3}$ for a strong light-matter interaction and a small photon loss $\gamma_c=10^{9}~\mathrm{s}^{-1}$, (ii)  a small $n_\mathrm{atom}=5\times10^{22}~\mathrm{m}^{-3}$ for a small light-matter interaction and a small photon loss $\gamma_c=10^{9}~\mathrm{s}^{-1}$, and (iii)  a large $n_\mathrm{atom}=5\times10^{24}~\mathrm{m}^{-3}$ for a strong light-matter interaction and a large photon loss $\gamma_c=5\times10^{12}~\mathrm{s}^{-1}$.

Figure ~\ref{fig:dy_com_eq} shows the dynamics of photons and electrons of the three cases. Regardless of the light-matter interaction and photon loss, all the electron populations are dominated by thermal relaxation. This results from the asymmetric coupling of light-matter interactions with atoms and photons. We define the two dimensionless coupling constants to the atoms and photons, respectively, from Eqs.~\eqref{eq:dne_f} and ~\eqref{eq:dNk_f} as 
\begin{align}
\tilde{g}_{LM}^{(a)} &= \frac{2\mu^2 n_V}{\hbar\epsilon_0\gamma},\\
\tilde{g}_{LM}^{(p)} &= \frac{2\mu^2 n_{\mathrm{atom}}}{\hbar\epsilon_0\gamma},
\end{align}
with the relation
\begin{align}
\tilde{g}_{LM}^{(p)} = N_j\tilde{g}_{LM}^{(a)}.
\end{align}
Typically,  $\tilde{g}_{LM}^{(a)}$ can be several orders of magnitude smaller than $\tilde{g}_{LM}^{(p)}$ since the number of the atoms $N_j$ in the volume $AL_c$ can be much more than one. In our simulation, we choose $N_j = 600$. Thus, the influence of light-matter interaction on electron dynamics is usually minor compared to thermal relaxation and pumping. In contrast, the interaction between light and matter can heavily influence photon dynamics due to a significant $\tilde{g}_{LM}^{(p)}$. 

Figures~\ref {fig:dy_com_eq}a and \ref{fig:dy_com_eq}d show the dynamics of photons and electrons in Case (i). The four frequencies cover the band-edge, band-center, and in-band-gap modes. As explained in the last paragraph, electron dynamics is dominated by thermal relaxation, approximated by
\begin{align}
n_n^{(e)}(t) \simeq f_n(T)\left(1-e^{-\gamma_r t}\right). \label{eq:ne_tr}
\end{align}
The populations in Fig.~\ref{fig:dy_com_eq}d exhibit an initial linear growth in time, followed by exponential saturation, which aligns well with Eq.~\eqref{eq:ne_tr}. Indeed, the temporal behavior described by Eq.~\eqref{eq:ne_tr} corresponds closely to all three cases, as illustrated in Figs.~\ref{fig:dy_com_eq}e and~\ref{fig:dy_com_eq}f for $\tilde{g}_{LM}^{(a)}$ is small.  Conversely, the behavior of photon dynamics shown in Fig.~\ref{fig:dy_com_eq}a varies significantly at different frequencies because of a strong coupling constant $\tilde{g}_{LM}^{(p)}=4.65$. The band-edge modes, which have greater confinement $\Gamma_k$, reach the steady states more quickly than the band-center and in-band-gap photons. The in-band-gap photons, with the smallest $\Gamma_k$, reach steady states in a time that is two orders of magnitude longer. However, the steady-state photon numbers for the four frequencies are similar, as described by the Bose-Einstein distribution. This shows that in Case (i), with strong light-matter interaction, band gaps primarily slow the dynamics of the photon number, while they do not suppress the steady-state photon numbers.

Figures~\ref {fig:dy_com_eq}b and \ref{fig:dy_com_eq}e show the dynamics of photons and electrons in Case (ii). In this case, light-matter interaction is much weaker, where the coupling constants $\tilde{g}_{LM}^{(a)}$ and $\tilde{g}_{LM}^{(p)}$ are reduced to one percent of their values in Case (i). Photons of all the frequencies in Fig.~\ref{fig:dy_com_eq}b show a slower dynamics compared to Fig.~\ref{fig:dy_com_eq}a.
The band-edge and band-center photons still reach the Bose-Einstein distribution. However, the in-band-gap photons reach smaller values than the Bose-Einstein distribution. This indicates that a weaker coupling strength of light-matter interaction leads to a reduction in photon numbers within the band gaps.

Figures~\ref {fig:dy_com_eq}c and \ref{fig:dy_com_eq}f show the dynamics of photons and electrons in Case (iii). In this case, the photon decay rate $\gamma_c =5\times 10^{12}~\mathrm{s}^{-1}$ is much larger than Case (i). Interestingly, the band-edge photons have similar dynamics to those in Case (i). Photons with high $\Gamma_k$ values appear to be unaffected by photon loss.  The band-center and in-band-gap photons saturate earlier at the same time scale, as compared to Fig.~\ref{fig:dy_com_eq}a. The band-center mode shows a slight reduction in steady-state photon numbers, while the in-band-gap mode exhibits a more significant decrease in these numbers.

The dynamics observed in the equilibrium regime can be summarized as follows: due to the relatively small contributions of both $\tilde{g}_{LM}^{(a)}$ and the pumping effects, thermal relaxation plays a dominant role in governing electron dynamics. The electron populations quickly saturate to the Fermi-Dirac distribution. Photons extract energy from electrons in a perturbative manner, characterized by a strength of $\tilde{g}_{LM}^{(p)}\Gamma_k$. Photons characterized by a large value of $\tilde{g}_{LM}^{(p)}\Gamma_k$, like those at the band edge, can effectively gain energy from electrons and achieve the Bose-Einstein distribution, even in the presence of a significant photon decay rate. In contrast, photons with a small value of $\tilde{g}_{LM}^{(p)}\Gamma_k$ are more susceptible to suppression due to band gaps and photon loss.

\begin{figure}
\includegraphics[width=6cm]{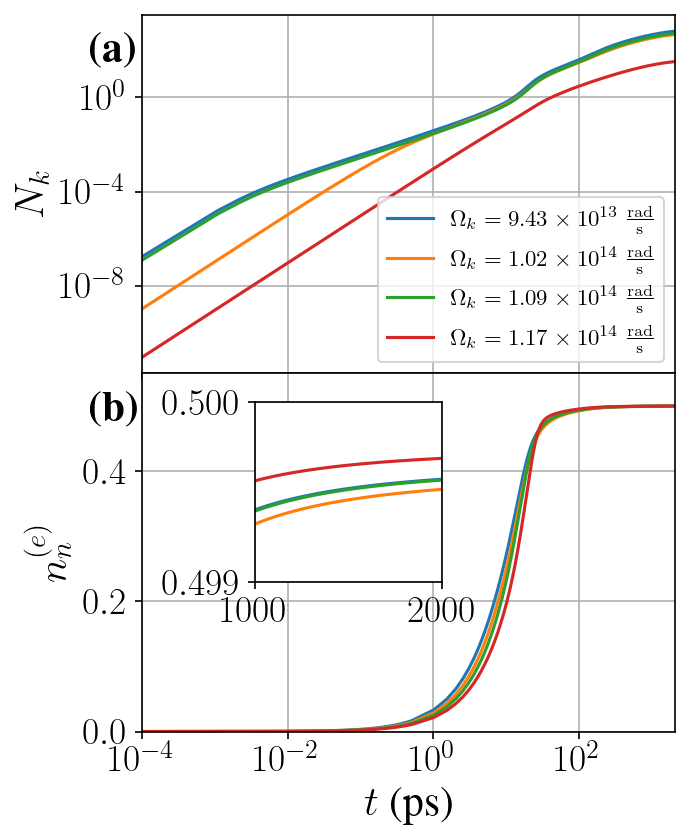}
\caption{\label{fig:dy_neq}
Dynamics of photons and electrons in the nonequilibrium regime with
 $\gamma_r = 10^{10}~\mathrm{s}^{-1}$,  $\Lambda_0= 10^{10}~\mathrm{s}^{-1}$, $\gamma_c=10^{9}~\mathrm{s}^{-1}$, and $n_\mathrm{atom}=5\times10^{24}~\mathrm{m}^{-3}$.  Panel (a) shows the multi-time-scale photon dynamics. Panel (b) shows the highly elevated electron populations, and the inset shows that the steady-state values are close to $\frac{1}{2}$. The electron populations $n^{(e)}_n(\omega)$ of $\omega = \Omega_k$ are plotted.  
}
\end{figure}

\subsection{Nonequilibrium}

For the nonequilibrium regime, we set the parameters:  $\gamma_r = 10^{10}~\mathrm{s}^{-1}$,  $\Lambda_0= 10^{10}~\mathrm{s}^{-1}$, $\gamma_c=10^{9}~\mathrm{s}^{-1}$, and $n_\mathrm{atom}=5\times10^{24}~\mathrm{m}^{-3}$. All the other parameters remain the same as in the equilibrium regime. In this regime, pumping is comparable to thermal relaxation, so that the steady-state electron populations deviate from the Fermi-Dirac distribution. Figure~\ref{fig:dy_neq} shows the multi-time-scale dynamics of photons and electrons in this regime. Since $\tilde{g}_{LM}^{(a)}\Gamma_k$ is still small, the electron populations are primarily influenced by pumping and thermal relaxation until $n^{(e)}_n(t)$ approaches a value of $\frac{1}{2}$. As $n^{(e)}_n(t) \simeq \frac{1}{2}$, strong stimulated emissions and absorptions occur to balance the pumping, keeping electron populations smaller than $\frac{1}{2}$. From Fig.~\ref{fig:dy_neq}b and its inset, it can be seen that the in-band-gap mode, which has a higher frequency, exhibits a higher steady-state electron population, while the other modes have lower steady-state electron populations. This observation contradicts the expected thermal distributions of both the reservoir and the pumping mechanisms. This phenomenon occurs because the band-edge modes and band-center modes possess greater confinement $\Gamma_k$, leading to increased stimulated emissions that help reduce electron populations. 

The roles of spontaneous emission and stimulated processes are evident in the photon dynamics in Fig.~\ref{fig:dy_neq}a. The photon numbers $N_k(t)$ of the band-edge modes show an initial quadratic growth in $t$, followed by linear growth, and then exponential saturation. 
The multi-stage process arises from competition among spontaneous emission, spontaneous absorption, and stimulated emission, particularly in cases of large confinement $\Gamma_k$. Initially, spontaneous emission dominates when $n^{(e)}_n$ and $N_k$ are small. However, as $n^{(e)}_n$ and $N_k$ increase, stimulated absorption becomes dominant first, followed by stimulated emission.
In contrast, the band-center mode and the in-band-gap mode, which have smaller $\Gamma_k$, display simpler dynamics due to weaker light-matter interaction. In the nonequilibrium regime, stimulated processes become more pronounced, causing the steady-state photon numbers to be more dependent on the band structures due to $\Gamma_k$. Such a dependence will be more evident in the discussion presented in Sec.~\ref{sec:ss}. In the nonequilibrium regime, it is important to note that numerical computations using the Runge-Kutta method can be time-consuming and may not be feasible for lower values of $\tilde{g}_{LM}^{(p)}$. Nevertheless, the steady-state solutions for lower values of $\tilde{g}_{LM}^{(p)}$ can still be obtained using the Newton-Krylov method. We present the steady-state solutions in the next Section.

\begin{figure*}
\includegraphics[width=12cm]{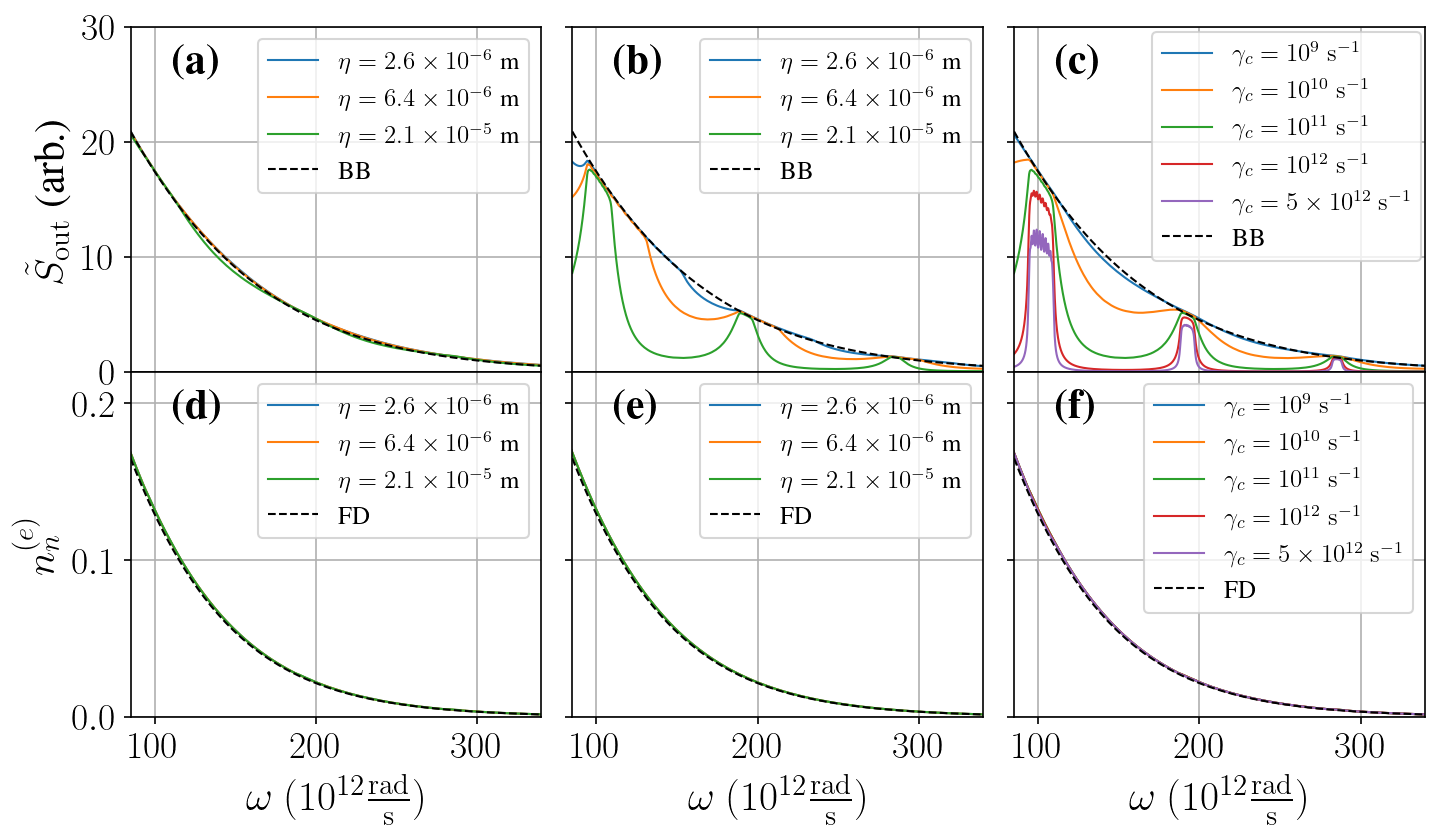}
\caption{\label{fig:ss_com_eq}
 Steady-state solutions of the emission spectra and the electron populations in the equilibrium regime. Panels (a) and (d) correspond to $n_\mathrm{atom}=5\times10^{24}~\mathrm{m}^{-3}$ and $\gamma_c =10^9~~\mathrm{s}^{-1}$.
Panels (b) and (e) correspond to $n_\mathrm{atom}=5\times10^{22}~\mathrm{m}^{-3}$ and $\gamma_c =10^9~~\mathrm{s}^{-1}$.
Panels (c) and (f) correspond to $n_\mathrm{atom}=5\times10^{24}~\mathrm{m}^{-3}$ and $\eta =2.1\times 10^{-5}~\mathrm{m}$. BB denotes the blackbody radiation. FD denotes the Fermi-Dirac distribution.
}
\end{figure*}

\section{Steady-State Solutions} \label{sec:ss}
We computed the steady-state solutions of the photon numbers of  Eqs.~\eqref{eq:dne_f} and ~\eqref{eq:dNk_f} by setting all the time derivatives to zero with the Newton-Krylov method. We utilize $N_k(t)$ and $n^{(e)}_n(t)$ from the Runge-Kutta method as initial guesses for the Newton-Krylov method, resulting in better convergence. The simulation parameters are listed in Table~\ref{tab:parameters}, except for those specified in each case.

We discuss the thermal emission spectrum $\tilde{S}_\mathrm{out}(\omega)$ in free space, defined in Eq.~\eqref{eq:Sout} for various  
$\eta$, $n_\mathrm{atom}$,   and $\gamma_c$ in the equilibrium regime ($\gamma_r=10^{13}~\mathrm{s}^{-1}$) and the nonequilibrium regime ($\gamma_r=10^{10}~\mathrm{s}^{-1}$). In both regimes, the primary factor influencing the thermal emission spectrum $\tilde{S}_\mathrm{out}(\omega)$ is the competition between light-matter interaction and photon decay. We will discuss how the parameters $\eta$, $n_{\text{atom}}$, and $\gamma_c$ influence the competition, as detailed below.

\subsection{Equilibrium}\label{subsec:ss_eq}

Figure~\ref{fig:ss_com_eq} shows the steady-state thermal emission spectrum $\tilde{S}_\mathrm{out}(\omega)$ and the steady-state electron populations in the equilibrium regime ($\gamma_r = 10^{13}~\mathrm{s}^{-1}$). Since thermal relaxation dominates the electron populations in all the cases,   
the electron populations $n^{(e)}_n$ follow closely the Fermi-Dirac distribution for all $\eta$, $n_{\text{atom}}$, and $\gamma_c$, as shown in Figs.~\ref{fig:ss_com_eq}d, \ref{fig:ss_com_eq}e, and \ref{fig:ss_com_eq}f. By varying $\eta$, the photonic band gaps are modified, as shown in Fig.~\ref{fig:band}. The band gaps appear to have no significant impact on the populations of electrons due to the small $g_{LM}^{(a)}$. In contrast, the band gaps can suppress the thermal emissions in some cases.

In Figs.~\ref{fig:ss_com_eq}a and~\ref{fig:ss_com_eq}d,  the parameters are $n_\mathrm{atom}=5\times10^{24}~\mathrm{m}^{-3}$ and $\gamma_c =10^9~~\mathrm{s}^{-1}$. In this case, light-matter interaction dominates, and $\gamma_c$ is negligible. Photons at all the frequencies can drain energy efficiently from electrons, so $\tilde{S}_{\mathrm{out}}$ for all $\eta$ can approach closely the blackbody radiation. The band gaps have negligible effects on $\tilde{S}_{\mathrm{out}}$.
 Note that  the blackbody radiation in one dimension is 
\begin{align}
\tilde{S}_{BB}(\omega) = \frac{\omega}{\exp\frac{\hbar\omega}{k_B T}-1}.
\end{align}

In Figs.~\ref{fig:ss_com_eq}b and~\ref{fig:ss_com_eq}e,  the parameters are $n_\mathrm{atom}=5\times10^{22}~\mathrm{m}^{-3}$ and $\gamma_c =10^9~~\mathrm{s}^{-1}$. 
In this case, light-matter interaction $\Gamma_k \tilde{g}_{LM}^{(p)}$ in the band gaps can be comparable to $\gamma_c$, resulting in a reduction of the thermal emission. Fig.~ \ref{fig:ss_com_eq}b shows that suppression of $\tilde{S}_{\mathrm{out}}(\omega)$ for different $\eta$ are characterized by the corresponding band gaps.

In Figs.~\ref{fig:ss_com_eq}c and~\ref{fig:ss_com_eq}f,  the parameters are $n_\mathrm{atom}=5\times10^{24}~\mathrm{m}^{-3}$ and $\eta = 2.1 \times 10^{-5}$~m. In this case, light-matter interaction is as strong as in Fig.~\ref{fig:ss_com_eq}a, and the band structure is depicted in Fig.~\ref{fig:band}c. As the photon decay rate increases, as shown in  Fig. ~\ref{fig:ss_com_eq}c, there is a noticeable suppression of thermal emission, particularly within the band gaps, in contrast to the nearly blackbody radiation in Fig.~\ref{fig:ss_com_eq}a. Increasing $\gamma_c$ renders the competition between light-matter interaction and photon loss appreciable, marking a regime where dissipation significantly alters the system dynamics.

In the equilibrium regime, thermal emission arises as photons passively acquire energy from electrons near the Fermi-Dirac distribution. For large $n_\mathrm{atom}$ and weak  $\gamma_c$, energy transfer is efficient across the spectrum, yielding Planckian emission. In contrast, when $n_\mathrm{atom}$ is reduced or $\gamma_c$ is increased, the competition between light–matter coupling and photon dissipation leads to pronounced deviations from equilibrium behavior, shaped by the underlying band structure.

\subsection{Nonequilibrium}
 \begin{figure}
\includegraphics[width=10cm]{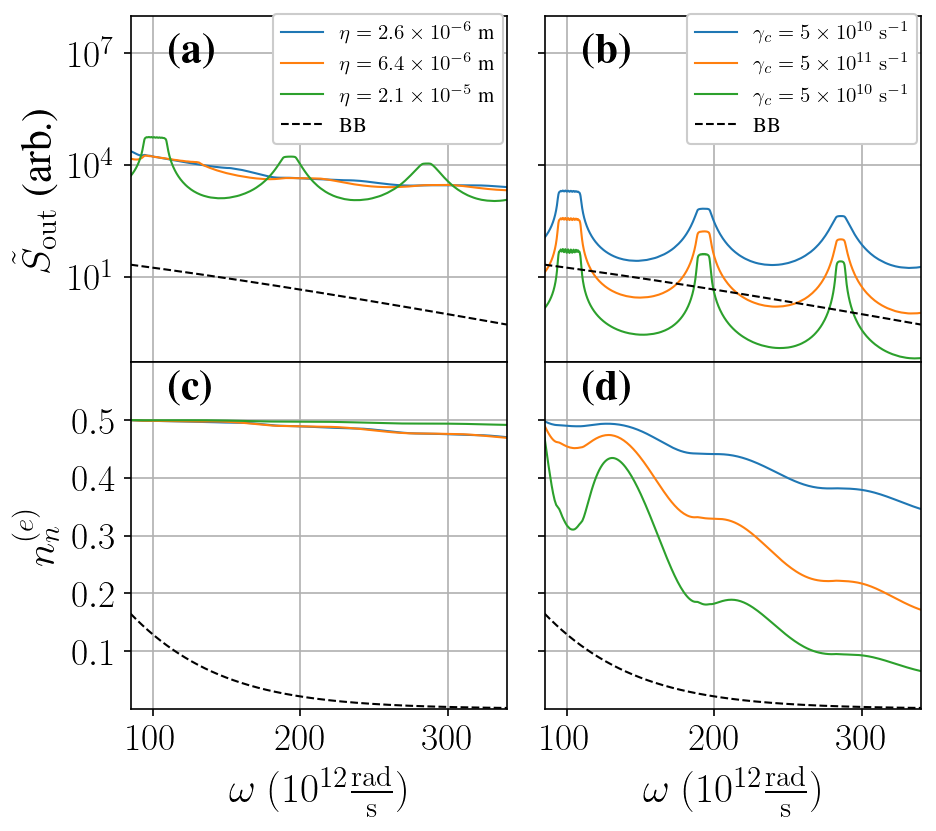}
\caption{ \label{fig:ss_com_neq}
Steady-state solutions of the emission spectra and the electron populations in the nonequilibrium regime. Panels (a) and (c) correspond to $n_\mathrm{atom}=5\times10^{24}~\mathrm{m}^{-3}$ and $\gamma_c =10^9~~\mathrm{s}^{-1}$.
Panels (b) and (d) correspond to $n_\mathrm{atom}=5\times10^{24}~\mathrm{m}^{-3}$ and $\eta =2.1\times10^{-5}~~\mathrm{m}$. 
The dashed lines in Panels (a) and (b) correspond to the blackbody radiation. The dashed lines in Panels (c) and (d) correspond to the Fermi-Dirac distribution.
}
\end{figure}

Figure~\ref{fig:ss_com_neq} presents the steady-state thermal emission spectrum $\tilde{S}_\mathrm{out}(\omega)$ and the corresponding electron populations in the nonequilibrium regime ($\gamma_r = 10^{10}~\mathrm{s}^{-1}$). Because the pumping strength $\Lambda_0$ becomes comparable to the thermal relaxation rate $\gamma_r$, the electron populations deviate markedly from the Fermi-Dirac distribution (Figs.~\ref{fig:ss_com_neq}c and \ref{fig:ss_com_neq}d), resulting in elevated population levels. Such high populations give rise to super-Planckian emission, with spectral intensity exceeding that of a blackbody at the same temperature. In this regime, stimulated processes become prominent for large $N_k$ and $n^{(e)}_n$, leading to nonperturbative light-matter interactions that modify both the thermal emission and the electron populations. We illustrate the nonperturbative light-matter interactions with the two cases: (i)  $n_\mathrm{atom}=5\times10^{24}~\mathrm{m}^{-3}$ and $\gamma_c =10^9~~\mathrm{s}^{-1}$ and (ii)  $n_\mathrm{atom}=5\times10^{24}~\mathrm{m}^{-3}$ and $\eta =2.1\times 10^{-5}~~\mathrm{m}$.

Figures~\ref{fig:ss_com_neq}a and \ref{fig:ss_com_neq}c show $\tilde{S}_{\mathrm{out}}(\omega)$ and $n^{(e)}_n(\omega)$ in Case~(i), where light-matter interaction dominates photon loss. The thermal emission spectra $\tilde{S}_\mathrm{out}(\omega)$ exceed the blackbody limit by three orders of magnitude. As $\eta$ varies, the thermal emission spectrum is modified and shaped by the underlying band structure, as shown in Fig.	~\ref{fig:ss_com_neq}a. Because of the large $N_k$,  the balance of strong stimulated absorption and emission leads $n^{(e)}_n(\omega)\lesssim\frac{1}{2}$ for the considered frequencies.

Figures~\ref{fig:ss_com_neq}b and \ref{fig:ss_com_neq}d show $\tilde{S}_{\mathrm{out}}(\omega)$ and $n^{(e)}_n(\omega)$ in Case~(ii), where the band structure is fixed with $\eta =2.1\times 10^{-5}~~\mathrm{m}$ and $\gamma_c$ varies.
As the photon decay rate $\gamma_c$ increases, photon loss competes with light-matter interaction, resulting in smaller $\tilde{S}_{\mathrm{out}}(\omega)$, as shown in Fig.~\ref{fig:ss_com_neq}b. All the thermal emission spectra are shaped by the band structure. Both the light-matter interaction and photon loss are strong, resulting in nonperturbative effects on the electron populations. As shown in Fig.~\ref{fig:ss_com_neq}d, $n^{(e)}_n(\omega)$ reflects a balance between stimulated processes, which drive the populations toward $\frac{1}{2}$, and photon loss, which suppresses them. Notably, in this intermediate regime, the electron distributions retain a clear imprint of the band structure as a result of the competition.

Taken together, these results demonstrate that thermal emission in the nonequilibrium regime is governed by a nonperturbative interplay between light–matter interaction and photon dissipation. In regimes where stimulated processes dominate, strong super-Planckian emission emerges with $n^{(e)}_n(\omega)\lesssim\frac{1}{2}$. Conversely, when photon loss becomes comparable, the system enters an intermediate regime in which both emission spectra and electron populations are jointly shaped by the band structure. The contrasting behaviors observed in Cases (i) and (ii) underscore the tunability of thermal emission via parameters such as $\gamma_c$ and $\eta$, offering a pathway to engineer emission spectra beyond conventional blackbody limits.

 \section{Concluding Remarks}\label{sec:conc}
 In conclusion, we have examined thermal emissions from a quantum photonic system with active pumping. Our findings reveal that variations in the strength of light-matter interaction and photon loss can lead to significantly different thermal emissions, even when thermal relaxation and pumping rates remain the same. The interplay between light-matter interaction strength and photon loss results in asynchronous and multi-time-scale dynamics of thermal emission, driven by spontaneous emission, stimulated processes, and photon dissipation.

We have discovered that in the equilibrium regime, the suppression of thermal emission due to band gaps can disappear when light-matter interaction becomes dominant. This observation suggests a new regime where band-gap suppression is absent, which contrasts with findings from earlier studies \cite{Cornelius1999, Lin2000, Lin2003, Narayanaswamy2004, Luo2004, han2007tailoring}. In the nonequilibrium regime, we find that the competition between light-matter interaction and photon loss plays a critical role in determining elevated electron populations and the degree of super-Planckian emission.

The engineering of thermal radiation has demonstrated promising applications in areas such as radiative cooling, thermal radiation control, and solar energy harvesting, capitalizing on the ability to manipulate thermal emissions in a passive manner \cite{fan2022photonics}. By uncovering the mechanisms governing thermal emission under active pumping and quantum two-level interactions, this work opens a route toward quantum control of thermal radiation.

\begin{acknowledgments}
We thank S. Y. Lin for the fruitful discussions. This work was supported by the National Science and Technology Council of Taiwan under Contract No. NSTC 114-2112-M-A49-025-.
\end{acknowledgments}

\appendix

\section{Simulation Parameters}
The simulation parameters used in this study are listed in Table~\ref{tab:parameters}. Simulations were done with the parameters in Table~\ref{tab:parameters} except those specified in the different cases. We assume that there are 600 atoms in the volume $A L_c$, which gives $n_\mathrm{atom} = \frac{600}{AL_c}$. It follows that $n_V = \frac{1}{AL_c}= \frac{n_\mathrm{atom}}{600}$. Note that we do not use  $A$ and $N_j$ directly as the input parameters for our simulations. We use $n_\mathrm{atom}$ and $n_\mathrm{V}$ as the input parameters, while maintaining the relation $n_\mathrm{atom} = N_j n_V$.

\begin{table}[h]
\caption{\label{tab:parameters}%
Simulation parameters.
}
\begin{ruledtabular}
\begin{tabular}{ccc}
\textrm{Symbol}&
\textrm{Meaning}&
\textrm{Value}\\
\colrule
$L$ & total system length & 1.2~cm\\
$L_c$ & photonic crystal length  & 120~$\mu$m\\
$l_p$ & lattice constant  & 10~$\mu$m\\
$n_V$& one over  the volume $AL_c$ & $n_{\mathrm{atom}}/600$\\
$\gamma$ & dephasing rate of polarization & $10^{14}~\mathrm{s}^{-1}$\\
$\gamma_c$ & photon decay rate & $10^{9}~\mathrm{s}^{-1}$\\
$\gamma_d$ & detector frequency resolution& $5\times10^{11}~\mathrm{s}^{-1}$\\
$\Lambda_0$ & pumping strength &$10^{10}~\mathrm{s}^{-1}$\\
$\omega_0$ & pumping  material band gap  &$1.6\times10^{14}~\frac{\mathrm{rad}}{\mathrm{s}}$\\
$\mu$ & dipole moment &$e\times1.3~$nm\\
$T$ & reservoir temperature& 400~K
\end{tabular}
\end{ruledtabular}
\end{table}


\bibliography{reference}

\providecommand{\noopsort}[1]{}\providecommand{\singleletter}[1]{#1}%
\begin{thebibliography}{25}%
\makeatletter
\providecommand \@ifxundefined [1]{%
 \@ifx{#1\undefined}
}%
\providecommand \@ifnum [1]{%
 \ifnum #1\expandafter \@firstoftwo
 \else \expandafter \@secondoftwo
 \fi
}%
\providecommand \@ifx [1]{%
 \ifx #1\expandafter \@firstoftwo
 \else \expandafter \@secondoftwo
 \fi
}%
\providecommand \natexlab [1]{#1}%
\providecommand \enquote  [1]{``#1''}%
\providecommand \bibnamefont  [1]{#1}%
\providecommand \bibfnamefont [1]{#1}%
\providecommand \citenamefont [1]{#1}%
\providecommand \href@noop [0]{\@secondoftwo}%
\providecommand \href [0]{\begingroup \@sanitize@url \@href}%
\providecommand \@href[1]{\@@startlink{#1}\@@href}%
\providecommand \@@href[1]{\endgroup#1\@@endlink}%
\providecommand \@sanitize@url [0]{\catcode `\\12\catcode `\$12\catcode
  `\&12\catcode `\#12\catcode `\^12\catcode `\_12\catcode `\%12\relax}%
\providecommand \@@startlink[1]{}%
\providecommand \@@endlink[0]{}%
\providecommand \url  [0]{\begingroup\@sanitize@url \@url }%
\providecommand \@url [1]{\endgroup\@href {#1}{\urlprefix }}%
\providecommand \urlprefix  [0]{URL }%
\providecommand \Eprint [0]{\href }%
\providecommand \doibase [0]{https://doi.org/}%
\providecommand \selectlanguage [0]{\@gobble}%
\providecommand \bibinfo  [0]{\@secondoftwo}%
\providecommand \bibfield  [0]{\@secondoftwo}%
\providecommand \translation [1]{[#1]}%
\providecommand \BibitemOpen [0]{}%
\providecommand \bibitemStop [0]{}%
\providecommand \bibitemNoStop [0]{.\EOS\space}%
\providecommand \EOS [0]{\spacefactor3000\relax}%
\providecommand \BibitemShut  [1]{\csname bibitem#1\endcsname}%
\let\auto@bib@innerbib\@empty
\bibitem [{\citenamefont {Landau}\ and\ \citenamefont
  {Lifshitz}(1980)}]{LandauLifshitz1980}%
  \BibitemOpen
  \bibfield  {author} {\bibinfo {author} {\bibfnamefont {L.~D.}\ \bibnamefont
  {Landau}}\ and\ \bibinfo {author} {\bibfnamefont {E.~M.}\ \bibnamefont
  {Lifshitz}},\ }\href@noop {} {\emph {\bibinfo {title} {Statistical
  Physics}}},\ \bibinfo {edition} {3rd}\ ed.,\ Vol.~\bibinfo {volume} {5}\
  (\bibinfo  {publisher} {Butterworth-Heinemann},\ \bibinfo {year}
  {1980})\BibitemShut {NoStop}%
\bibitem [{\citenamefont {Baranov}\ \emph {et~al.}(2019)\citenamefont
  {Baranov}, \citenamefont {Xiao}, \citenamefont {Nechepurenko}, \citenamefont
  {Krasnok}, \citenamefont {Al{\`u}},\ and\ \citenamefont
  {Kats}}]{baranov2019nanophotonic}%
  \BibitemOpen
  \bibfield  {author} {\bibinfo {author} {\bibfnamefont {D.~G.}\ \bibnamefont
  {Baranov}}, \bibinfo {author} {\bibfnamefont {Y.}~\bibnamefont {Xiao}},
  \bibinfo {author} {\bibfnamefont {I.~A.}\ \bibnamefont {Nechepurenko}},
  \bibinfo {author} {\bibfnamefont {A.}~\bibnamefont {Krasnok}}, \bibinfo
  {author} {\bibfnamefont {A.}~\bibnamefont {Al{\`u}}},\ and\ \bibinfo {author}
  {\bibfnamefont {M.~A.}\ \bibnamefont {Kats}},\ }\bibfield  {title} {\bibinfo
  {title} {Nanophotonic engineering of far-field thermal emitters},\
  }\href@noop {} {\bibfield  {journal} {\bibinfo  {journal} {Nat. Mater.}\
  }\textbf {\bibinfo {volume} {18}},\ \bibinfo {pages} {920} (\bibinfo {year}
  {2019})}\BibitemShut {NoStop}%
\bibitem [{\citenamefont {Cornelius}\ and\ \citenamefont
  {Dowling}(1999)}]{Cornelius1999}%
  \BibitemOpen
  \bibfield  {author} {\bibinfo {author} {\bibfnamefont {C.~M.}\ \bibnamefont
  {Cornelius}}\ and\ \bibinfo {author} {\bibfnamefont {J.~P.}\ \bibnamefont
  {Dowling}},\ }\bibfield  {title} {\bibinfo {title} {Modification of planck
  blackbody radiation by photonic band-gap structures},\ }\href
  {https://doi.org/10.1103/PhysRevA.59.4736} {\bibfield  {journal} {\bibinfo
  {journal} {Phys. Rev. A}\ }\textbf {\bibinfo {volume} {59}},\ \bibinfo
  {pages} {4736} (\bibinfo {year} {1999})}\BibitemShut {NoStop}%
\bibitem [{\citenamefont {Lin}\ \emph {et~al.}(2000)\citenamefont {Lin},
  \citenamefont {Fleming}, \citenamefont {Chow}, \citenamefont {Bur},
  \citenamefont {Choi},\ and\ \citenamefont {Goldberg}}]{Lin2000}%
  \BibitemOpen
  \bibfield  {author} {\bibinfo {author} {\bibfnamefont {S.-Y.}\ \bibnamefont
  {Lin}}, \bibinfo {author} {\bibfnamefont {J.~G.}\ \bibnamefont {Fleming}},
  \bibinfo {author} {\bibfnamefont {E.}~\bibnamefont {Chow}}, \bibinfo {author}
  {\bibfnamefont {J.}~\bibnamefont {Bur}}, \bibinfo {author} {\bibfnamefont
  {K.~K.}\ \bibnamefont {Choi}},\ and\ \bibinfo {author} {\bibfnamefont
  {A.}~\bibnamefont {Goldberg}},\ }\bibfield  {title} {\bibinfo {title}
  {Enhancement and suppression of thermal emission by a three-dimensional
  photonic crystal},\ }\href {https://doi.org/10.1103/PhysRevB.62.R2243}
  {\bibfield  {journal} {\bibinfo  {journal} {Phys. Rev. B}\ }\textbf {\bibinfo
  {volume} {62}},\ \bibinfo {pages} {R2243} (\bibinfo {year}
  {2000})}\BibitemShut {NoStop}%
\bibitem [{\citenamefont {Lin}\ \emph {et~al.}(2003)\citenamefont {Lin},
  \citenamefont {Moreno},\ and\ \citenamefont {Fleming}}]{Lin2003}%
  \BibitemOpen
  \bibfield  {author} {\bibinfo {author} {\bibfnamefont {S.~Y.}\ \bibnamefont
  {Lin}}, \bibinfo {author} {\bibfnamefont {J.}~\bibnamefont {Moreno}},\ and\
  \bibinfo {author} {\bibfnamefont {J.~G.}\ \bibnamefont {Fleming}},\
  }\bibfield  {title} {\bibinfo {title} {Three-dimensional photonic-crystal
  emitter for thermal photovoltaic power generation},\ }\href
  {https://doi.org/10.1063/1.1592614} {\bibfield  {journal} {\bibinfo
  {journal} {Appl. Phys. Lett.}\ }\textbf {\bibinfo {volume} {83}},\ \bibinfo
  {pages} {380} (\bibinfo {year} {2003})}\BibitemShut {NoStop}%
\bibitem [{\citenamefont {Narayanaswamy}\ and\ \citenamefont
  {Chen}(2004)}]{Narayanaswamy2004}%
  \BibitemOpen
  \bibfield  {author} {\bibinfo {author} {\bibfnamefont {A.}~\bibnamefont
  {Narayanaswamy}}\ and\ \bibinfo {author} {\bibfnamefont {G.}~\bibnamefont
  {Chen}},\ }\bibfield  {title} {\bibinfo {title} {Thermal emission control
  with one-dimensional metallodielectric photonic crystals},\ }\href
  {https://doi.org/10.1103/PhysRevB.70.125101} {\bibfield  {journal} {\bibinfo
  {journal} {Phys. Rev. B}\ }\textbf {\bibinfo {volume} {70}},\ \bibinfo
  {pages} {125101} (\bibinfo {year} {2004})}\BibitemShut {NoStop}%
\bibitem [{\citenamefont {Luo}\ \emph {et~al.}(2004)\citenamefont {Luo},
  \citenamefont {Narayanaswamy}, \citenamefont {Chen},\ and\ \citenamefont
  {Joannopoulos}}]{Luo2004}%
  \BibitemOpen
  \bibfield  {author} {\bibinfo {author} {\bibfnamefont {C.}~\bibnamefont
  {Luo}}, \bibinfo {author} {\bibfnamefont {A.}~\bibnamefont {Narayanaswamy}},
  \bibinfo {author} {\bibfnamefont {G.}~\bibnamefont {Chen}},\ and\ \bibinfo
  {author} {\bibfnamefont {J.~D.}\ \bibnamefont {Joannopoulos}},\ }\bibfield
  {title} {\bibinfo {title} {Thermal radiation from photonic crystals: A direct
  calculation},\ }\href {https://doi.org/10.1103/PhysRevLett.93.213905}
  {\bibfield  {journal} {\bibinfo  {journal} {Phys. Rev. Lett.}\ }\textbf
  {\bibinfo {volume} {93}},\ \bibinfo {pages} {213905} (\bibinfo {year}
  {2004})}\BibitemShut {NoStop}%
\bibitem [{\citenamefont {Han}\ \emph {et~al.}(2007)\citenamefont {Han},
  \citenamefont {Stein},\ and\ \citenamefont {Norris}}]{han2007tailoring}%
  \BibitemOpen
  \bibfield  {author} {\bibinfo {author} {\bibfnamefont {S.~E.}\ \bibnamefont
  {Han}}, \bibinfo {author} {\bibfnamefont {A.}~\bibnamefont {Stein}},\ and\
  \bibinfo {author} {\bibfnamefont {D.~J.}\ \bibnamefont {Norris}},\ }\bibfield
   {title} {\bibinfo {title} {Tailoring self-assembled metallic photonic
  crystals for modified thermal emission},\ }\href@noop {} {\bibfield
  {journal} {\bibinfo  {journal} {Phys. Rev. Lett.}\ }\textbf {\bibinfo
  {volume} {99}},\ \bibinfo {pages} {053906} (\bibinfo {year}
  {2007})}\BibitemShut {NoStop}%
\bibitem [{\citenamefont {Kravets}\ \emph {et~al.}(2008)\citenamefont
  {Kravets}, \citenamefont {Schedin},\ and\ \citenamefont
  {Grigorenko}}]{kravets2008plasmonic}%
  \BibitemOpen
  \bibfield  {author} {\bibinfo {author} {\bibfnamefont {V.}~\bibnamefont
  {Kravets}}, \bibinfo {author} {\bibfnamefont {F.}~\bibnamefont {Schedin}},\
  and\ \bibinfo {author} {\bibfnamefont {A.}~\bibnamefont {Grigorenko}},\
  }\bibfield  {title} {\bibinfo {title} {Plasmonic blackbody: Almost complete
  absorption of light in nanostructured metallic coatings},\ }\href@noop {}
  {\bibfield  {journal} {\bibinfo  {journal} {Phys. Rev. B}\ }\textbf {\bibinfo
  {volume} {78}},\ \bibinfo {pages} {205405} (\bibinfo {year}
  {2008})}\BibitemShut {NoStop}%
\bibitem [{\citenamefont {Costantini}\ \emph {et~al.}(2015)\citenamefont
  {Costantini}, \citenamefont {Lefebvre}, \citenamefont {Coutrot},
  \citenamefont {Moldovan-Doyen}, \citenamefont {Hugonin}, \citenamefont
  {Boutami}, \citenamefont {Marquier}, \citenamefont {Benisty},\ and\
  \citenamefont {Greffet}}]{Costantini2015}%
  \BibitemOpen
  \bibfield  {author} {\bibinfo {author} {\bibfnamefont {D.}~\bibnamefont
  {Costantini}}, \bibinfo {author} {\bibfnamefont {A.}~\bibnamefont
  {Lefebvre}}, \bibinfo {author} {\bibfnamefont {A.-L.}\ \bibnamefont
  {Coutrot}}, \bibinfo {author} {\bibfnamefont {I.}~\bibnamefont
  {Moldovan-Doyen}}, \bibinfo {author} {\bibfnamefont {J.-P.}\ \bibnamefont
  {Hugonin}}, \bibinfo {author} {\bibfnamefont {S.}~\bibnamefont {Boutami}},
  \bibinfo {author} {\bibfnamefont {F.}~\bibnamefont {Marquier}}, \bibinfo
  {author} {\bibfnamefont {H.}~\bibnamefont {Benisty}},\ and\ \bibinfo {author}
  {\bibfnamefont {J.-J.}\ \bibnamefont {Greffet}},\ }\bibfield  {title}
  {\bibinfo {title} {Plasmonic metasurface for directional and
  frequency-selective thermal emission},\ }\href
  {https://doi.org/10.1103/PhysRevApplied.4.014023} {\bibfield  {journal}
  {\bibinfo  {journal} {Phys. Rev. Appl.}\ }\textbf {\bibinfo {volume} {4}},\
  \bibinfo {pages} {014023} (\bibinfo {year} {2015})}\BibitemShut {NoStop}%
\bibitem [{\citenamefont {Aydin}\ \emph {et~al.}(2011)\citenamefont {Aydin},
  \citenamefont {Ferry}, \citenamefont {Briggs},\ and\ \citenamefont
  {Atwater}}]{aydin2011broadband}%
  \BibitemOpen
  \bibfield  {author} {\bibinfo {author} {\bibfnamefont {K.}~\bibnamefont
  {Aydin}}, \bibinfo {author} {\bibfnamefont {V.~E.}\ \bibnamefont {Ferry}},
  \bibinfo {author} {\bibfnamefont {R.~M.}\ \bibnamefont {Briggs}},\ and\
  \bibinfo {author} {\bibfnamefont {H.~A.}\ \bibnamefont {Atwater}},\
  }\bibfield  {title} {\bibinfo {title} {Broadband polarization-independent
  resonant light absorption using ultrathin plasmonic super absorbers},\
  }\href@noop {} {\bibfield  {journal} {\bibinfo  {journal} {Nat. Commun.}\
  }\textbf {\bibinfo {volume} {2}},\ \bibinfo {pages} {517} (\bibinfo {year}
  {2011})}\BibitemShut {NoStop}%
\bibitem [{\citenamefont {Kittel}\ \emph {et~al.}(2005)\citenamefont {Kittel},
  \citenamefont {M\"uller-Hirsch}, \citenamefont {Parisi}, \citenamefont
  {Biehs}, \citenamefont {Reddig},\ and\ \citenamefont
  {Holthaus}}]{Kittel2005}%
  \BibitemOpen
  \bibfield  {author} {\bibinfo {author} {\bibfnamefont {A.}~\bibnamefont
  {Kittel}}, \bibinfo {author} {\bibfnamefont {W.}~\bibnamefont
  {M\"uller-Hirsch}}, \bibinfo {author} {\bibfnamefont {J.}~\bibnamefont
  {Parisi}}, \bibinfo {author} {\bibfnamefont {S.-A.}\ \bibnamefont {Biehs}},
  \bibinfo {author} {\bibfnamefont {D.}~\bibnamefont {Reddig}},\ and\ \bibinfo
  {author} {\bibfnamefont {M.}~\bibnamefont {Holthaus}},\ }\bibfield  {title}
  {\bibinfo {title} {Near-field heat transfer in a scanning thermal
  microscope},\ }\href {https://doi.org/10.1103/PhysRevLett.95.224301}
  {\bibfield  {journal} {\bibinfo  {journal} {Phys. Rev. Lett.}\ }\textbf
  {\bibinfo {volume} {95}},\ \bibinfo {pages} {224301} (\bibinfo {year}
  {2005})}\BibitemShut {NoStop}%
\bibitem [{\citenamefont {Ben-Abdallah}\ \emph {et~al.}(2008)\citenamefont
  {Ben-Abdallah}, \citenamefont {Joulain}, \citenamefont {Drevillon},\ and\
  \citenamefont {Le~Goff}}]{Ben-Abdallah2008}%
  \BibitemOpen
  \bibfield  {author} {\bibinfo {author} {\bibfnamefont {P.}~\bibnamefont
  {Ben-Abdallah}}, \bibinfo {author} {\bibfnamefont {K.}~\bibnamefont
  {Joulain}}, \bibinfo {author} {\bibfnamefont {J.}~\bibnamefont {Drevillon}},\
  and\ \bibinfo {author} {\bibfnamefont {C.}~\bibnamefont {Le~Goff}},\
  }\bibfield  {title} {\bibinfo {title} {Heat transport through plasmonic
  interactions in closely spaced metallic nanoparticle chains},\ }\href
  {https://doi.org/10.1103/PhysRevB.77.075417} {\bibfield  {journal} {\bibinfo
  {journal} {Phys. Rev. B}\ }\textbf {\bibinfo {volume} {77}},\ \bibinfo
  {pages} {075417} (\bibinfo {year} {2008})}\BibitemShut {NoStop}%
\bibitem [{\citenamefont {Nolen}\ \emph {et~al.}(2024)\citenamefont {Nolen},
  \citenamefont {Overvig}, \citenamefont {Cotrufo},\ and\ \citenamefont
  {Al{\`u}}}]{nolen2024local}%
  \BibitemOpen
  \bibfield  {author} {\bibinfo {author} {\bibfnamefont {J.~R.}\ \bibnamefont
  {Nolen}}, \bibinfo {author} {\bibfnamefont {A.~C.}\ \bibnamefont {Overvig}},
  \bibinfo {author} {\bibfnamefont {M.}~\bibnamefont {Cotrufo}},\ and\ \bibinfo
  {author} {\bibfnamefont {A.}~\bibnamefont {Al{\`u}}},\ }\bibfield  {title}
  {\bibinfo {title} {Local control of polarization and geometric phase in
  thermal metasurfaces},\ }\href@noop {} {\bibfield  {journal} {\bibinfo
  {journal} {Nat. Nanotechnol.}\ }\textbf {\bibinfo {volume} {19}},\ \bibinfo
  {pages} {1627} (\bibinfo {year} {2024})}\BibitemShut {NoStop}%
\bibitem [{\citenamefont {Li}\ and\ \citenamefont
  {Fan}(2018)}]{li2018nanophotonic}%
  \BibitemOpen
  \bibfield  {author} {\bibinfo {author} {\bibfnamefont {W.}~\bibnamefont
  {Li}}\ and\ \bibinfo {author} {\bibfnamefont {S.}~\bibnamefont {Fan}},\
  }\bibfield  {title} {\bibinfo {title} {Nanophotonic control of thermal
  radiation for energy applications},\ }\href@noop {} {\bibfield  {journal}
  {\bibinfo  {journal} {Opt. Express}\ }\textbf {\bibinfo {volume} {26}},\
  \bibinfo {pages} {15995} (\bibinfo {year} {2018})}\BibitemShut {NoStop}%
\bibitem [{\citenamefont {Wurfel}(1982)}]{wurfel1982chemical}%
  \BibitemOpen
  \bibfield  {author} {\bibinfo {author} {\bibfnamefont {P.}~\bibnamefont
  {Wurfel}},\ }\bibfield  {title} {\bibinfo {title} {The chemical potential of
  radiation},\ }\href@noop {} {\bibfield  {journal} {\bibinfo  {journal} {J.
  Phys. C: Solid State Phys.}\ }\textbf {\bibinfo {volume} {15}},\ \bibinfo
  {pages} {3967} (\bibinfo {year} {1982})}\BibitemShut {NoStop}%
\bibitem [{\citenamefont {Hsieh}\ \emph {et~al.}(2025)\citenamefont {Hsieh},
  \citenamefont {Romero}, \citenamefont {Bur}, \citenamefont {Wang},
  \citenamefont {Wu}, \citenamefont {Narayan}, \citenamefont {John},\ and\
  \citenamefont {Lin}}]{hsieh2025observation}%
  \BibitemOpen
  \bibfield  {author} {\bibinfo {author} {\bibfnamefont {M.-L.}\ \bibnamefont
  {Hsieh}}, \bibinfo {author} {\bibfnamefont {T.}~\bibnamefont {Romero}},
  \bibinfo {author} {\bibfnamefont {J.}~\bibnamefont {Bur}}, \bibinfo {author}
  {\bibfnamefont {X.}~\bibnamefont {Wang}}, \bibinfo {author} {\bibfnamefont
  {J.-S.}\ \bibnamefont {Wu}}, \bibinfo {author} {\bibfnamefont
  {S.}~\bibnamefont {Narayan}}, \bibinfo {author} {\bibfnamefont
  {S.}~\bibnamefont {John}},\ and\ \bibinfo {author} {\bibfnamefont {S.-Y.}\
  \bibnamefont {Lin}},\ }\bibfield  {title} {\bibinfo {title} {Observation of
  nonlinear input-output power characteristic of light emission from an
  electrically-biased metallic photonic crystal},\ }\href@noop {} {\bibfield
  {journal} {\bibinfo  {journal} {npj Nanophotonics}\ }\textbf {\bibinfo
  {volume} {2}},\ \bibinfo {pages} {15} (\bibinfo {year} {2025})}\BibitemShut
  {NoStop}%
\bibitem [{\citenamefont {Ridolfo}\ \emph {et~al.}(2013)\citenamefont
  {Ridolfo}, \citenamefont {Savasta},\ and\ \citenamefont
  {Hartmann}}]{ridolfo2013nonclassical}%
  \BibitemOpen
  \bibfield  {author} {\bibinfo {author} {\bibfnamefont {A.}~\bibnamefont
  {Ridolfo}}, \bibinfo {author} {\bibfnamefont {S.}~\bibnamefont {Savasta}},\
  and\ \bibinfo {author} {\bibfnamefont {M.~J.}\ \bibnamefont {Hartmann}},\
  }\bibfield  {title} {\bibinfo {title} {Nonclassical radiation from thermal
  cavities in the ultrastrong coupling regime},\ }\href@noop {} {\bibfield
  {journal} {\bibinfo  {journal} {Phys. Rev. Lett.}\ }\textbf {\bibinfo
  {volume} {110}},\ \bibinfo {pages} {163601} (\bibinfo {year}
  {2013})}\BibitemShut {NoStop}%
\bibitem [{\citenamefont {Pirkkalainen}\ \emph {et~al.}(2015)\citenamefont
  {Pirkkalainen}, \citenamefont {Cho}, \citenamefont {Massel}, \citenamefont
  {Tuorila}, \citenamefont {Heikkil{\"a}}, \citenamefont {Hakonen},\ and\
  \citenamefont {Sillanp{\"a}{\"a}}}]{pirkkalainen2015cavity}%
  \BibitemOpen
  \bibfield  {author} {\bibinfo {author} {\bibfnamefont {J.-M.}\ \bibnamefont
  {Pirkkalainen}}, \bibinfo {author} {\bibfnamefont {S.}~\bibnamefont {Cho}},
  \bibinfo {author} {\bibfnamefont {F.}~\bibnamefont {Massel}}, \bibinfo
  {author} {\bibfnamefont {J.}~\bibnamefont {Tuorila}}, \bibinfo {author}
  {\bibfnamefont {T.}~\bibnamefont {Heikkil{\"a}}}, \bibinfo {author}
  {\bibfnamefont {P.}~\bibnamefont {Hakonen}},\ and\ \bibinfo {author}
  {\bibfnamefont {M.}~\bibnamefont {Sillanp{\"a}{\"a}}},\ }\bibfield  {title}
  {\bibinfo {title} {Cavity optomechanics mediated by a quantum two-level
  system},\ }\href@noop {} {\bibfield  {journal} {\bibinfo  {journal} {Nat.
  Commun.}\ }\textbf {\bibinfo {volume} {6}},\ \bibinfo {pages} {6981}
  (\bibinfo {year} {2015})}\BibitemShut {NoStop}%
\bibitem [{\citenamefont {Askenazi}\ \emph {et~al.}(2017)\citenamefont
  {Askenazi}, \citenamefont {Vasanelli}, \citenamefont {Todorov}, \citenamefont
  {Sakat}, \citenamefont {Greffet}, \citenamefont {Beaudoin}, \citenamefont
  {Sagnes},\ and\ \citenamefont {Sirtori}}]{askenazi2017midinfrared}%
  \BibitemOpen
  \bibfield  {author} {\bibinfo {author} {\bibfnamefont {B.}~\bibnamefont
  {Askenazi}}, \bibinfo {author} {\bibfnamefont {A.}~\bibnamefont {Vasanelli}},
  \bibinfo {author} {\bibfnamefont {Y.}~\bibnamefont {Todorov}}, \bibinfo
  {author} {\bibfnamefont {E.}~\bibnamefont {Sakat}}, \bibinfo {author}
  {\bibfnamefont {J.-J.}\ \bibnamefont {Greffet}}, \bibinfo {author}
  {\bibfnamefont {G.}~\bibnamefont {Beaudoin}}, \bibinfo {author}
  {\bibfnamefont {I.}~\bibnamefont {Sagnes}},\ and\ \bibinfo {author}
  {\bibfnamefont {C.}~\bibnamefont {Sirtori}},\ }\bibfield  {title} {\bibinfo
  {title} {Midinfrared ultrastrong light--matter coupling for {THz} thermal
  emission},\ }\href@noop {} {\bibfield  {journal} {\bibinfo  {journal} {ACS
  Photonics}\ }\textbf {\bibinfo {volume} {4}},\ \bibinfo {pages} {2550}
  (\bibinfo {year} {2017})}\BibitemShut {NoStop}%
\bibitem [{\citenamefont {Pilar}\ \emph {et~al.}(2020)\citenamefont {Pilar},
  \citenamefont {De~Bernardis},\ and\ \citenamefont
  {Rabl}}]{Pilar2020thermodynamicsof}%
  \BibitemOpen
  \bibfield  {author} {\bibinfo {author} {\bibfnamefont {P.}~\bibnamefont
  {Pilar}}, \bibinfo {author} {\bibfnamefont {D.}~\bibnamefont
  {De~Bernardis}},\ and\ \bibinfo {author} {\bibfnamefont {P.}~\bibnamefont
  {Rabl}},\ }\bibfield  {title} {\bibinfo {title} {Thermodynamics of
  ultrastrongly coupled light-matter systems},\ }\href
  {https://doi.org/10.22331/q-2020-09-28-335} {\bibfield  {journal} {\bibinfo
  {journal} {{Quantum}}\ }\textbf {\bibinfo {volume} {4}},\ \bibinfo {pages}
  {335} (\bibinfo {year} {2020})}\BibitemShut {NoStop}%
\bibitem [{\citenamefont {Chow}(2006)}]{chow2006theory}%
  \BibitemOpen
  \bibfield  {author} {\bibinfo {author} {\bibfnamefont {W.~W.}\ \bibnamefont
  {Chow}},\ }\bibfield  {title} {\bibinfo {title} {Theory of emission from an
  active photonic lattice},\ }\href@noop {} {\bibfield  {journal} {\bibinfo
  {journal} {Phys. Rev. A}\ }\textbf {\bibinfo {volume} {73}},\ \bibinfo
  {pages} {013821} (\bibinfo {year} {2006})}\BibitemShut {NoStop}%
\bibitem [{\citenamefont {Lang}\ \emph {et~al.}(1973)\citenamefont {Lang},
  \citenamefont {Scully},\ and\ \citenamefont {Lamb}}]{Lang1973}%
  \BibitemOpen
  \bibfield  {author} {\bibinfo {author} {\bibfnamefont {R.}~\bibnamefont
  {Lang}}, \bibinfo {author} {\bibfnamefont {M.~O.}\ \bibnamefont {Scully}},\
  and\ \bibinfo {author} {\bibfnamefont {W.~E.}\ \bibnamefont {Lamb}},\
  }\bibfield  {title} {\bibinfo {title} {Why is the laser line so narrow? a
  theory of single-quasimode laser operation},\ }\href
  {https://doi.org/10.1103/PhysRevA.7.1788} {\bibfield  {journal} {\bibinfo
  {journal} {Phys. Rev. A}\ }\textbf {\bibinfo {volume} {7}},\ \bibinfo {pages}
  {1788} (\bibinfo {year} {1973})}\BibitemShut {NoStop}%
\bibitem [{\citenamefont {Gerry}\ and\ \citenamefont
  {Knight}(2023)}]{gerry2023introductory}%
  \BibitemOpen
  \bibfield  {author} {\bibinfo {author} {\bibfnamefont {C.~C.}\ \bibnamefont
  {Gerry}}\ and\ \bibinfo {author} {\bibfnamefont {P.~L.}\ \bibnamefont
  {Knight}},\ }\href@noop {} {\emph {\bibinfo {title} {Introductory quantum
  optics}}}\ (\bibinfo  {publisher} {Cambridge University Press},\ \bibinfo
  {year} {2023})\BibitemShut {NoStop}%
\bibitem [{\citenamefont {Fan}\ and\ \citenamefont
  {Li}(2022)}]{fan2022photonics}%
  \BibitemOpen
  \bibfield  {author} {\bibinfo {author} {\bibfnamefont {S.}~\bibnamefont
  {Fan}}\ and\ \bibinfo {author} {\bibfnamefont {W.}~\bibnamefont {Li}},\
  }\bibfield  {title} {\bibinfo {title} {Photonics and thermodynamics concepts
  in radiative cooling},\ }\href@noop {} {\bibfield  {journal} {\bibinfo
  {journal} {Nat. Photonics}\ }\textbf {\bibinfo {volume} {16}},\ \bibinfo
  {pages} {182} (\bibinfo {year} {2022})}\BibitemShut {NoStop}%
\end{thebibliography}%

\end{document}